\documentclass{pas}
\usepackage[nopatch]{microtype}
\usepackage{booktabs, multirow, changepage}
\usepackage{xspace}
\usepackage[table]{xcolor}
\usepackage{ragged2e}
\usepackage{array, rotating}
\usepackage{graphicx}
\usepackage{longtable}
\usepackage{booktabs}
\usepackage{threeparttablex}
\usepackage{subcaption}
\usepackage{makecell}
\newcolumntype{L}[1]{>{\RaggedRight\arraybackslash}p{#1}}
\definecolor{headerblue}{HTML}{0023a0}
\definecolor{lightblue}{HTML}{e2e8ff}
\newcommand{\qmist}{\texttt{QMIST}\xspace}
\begin{document}
\lefttitle{Publications of the Astronomical Society of Australia}
\righttitle{Amarnath and Y. Maan}
\jnlPage{1}{4}
\jnlDoiYr{2021}
\doival{10.1017/pasa.xxxx.xx}
\articletitt{Research Paper}
\title{Quasi-Periodic Microstructures in Pulsar Emission: Automated Detection and Archival Survey}

\author{\gn{Amarnath}$^{1,2,3}$ and \gn{Yogesh} \sn{Maan}$^{1}$}
\affil{$^1$National Centre for Radio Astrophysics, Pune 411007, Maharashtra, India}
\affil{$^2$Cochin University of Science and Technology, Kochi 682022, Kerala, India}
\affil{$^3$Raman Research Institute, Bangalore 560080, Karnataka, India}
\corresp{Yogesh Maan, Email: ymaan@ncra.tifr.res.in}
\begin{abstract}
The study of quasi-period microstructures in pulsars offers valuable insights into the underlying emission mechanism. However, identifying these features through manual inspection of the intensity time series, often containing thousands to millions of pulses, is both laborious and time-consuming. To address this challenge, we have developed a Python-based software, Quasi-periodic MIcrostructure Search Tool (\qmist), to automate the search for quasi-periodic microstructures in radio pulsar time-series data. We provide a detailed description of the algorithms used in \qmist, demonstrate its efficacy using data on pulsars known to exhibit microstructures, and discuss potential future improvements. Using \qmist, we have performed a multi-epoch survey of quasi-periodic microstructures in a sample of 27 pulsars, using observations from the Giant Metrewave Radio Telescope and the Green Bank Telescope, as well as the archival data from the Parkes telescope. In addition to recovering previously reported microstructures from several pulsars, we report, for the first time, detection of quasi-periodic microstructures in three pulsars, B1451$-$68, B1706$-$16 and B1845$-$19. We also estimate the typical period of microstructures in another pulsar, B0540+23, that was known to exhibit microstructures earlier, but the periodicity was unknown. Using the periodicity measurements from our survey, and earlier such measurements from the literature, we confirm the near linear relationship between the microstructure periodicity and the rotation period of pulsars, and discuss our results in the context of the emission mechanism of microstructures.
\end{abstract}
\maketitle
\section{Introduction}
The emission mechanism of pulsars is not entirely understood even after several decades of observations. The radio emission from these rapidly rotating neutron stars exhibits a variety of phenomena across a wide range of timescales. One of the features of pulsar emission, called microstructures, is often observed as a narrow train of quasi-periodic peaks on top of a fraction of single pulses. These short time-scale structures, first discovered by \citet{craft1968submillisecond}, have been detected in several pulsars as revealed in numerous studies such as \citet{kramer2002high} in Vela pulsar, \citet{lange1998radio} in B1133+16, and many others. \citet{mitra2015polarized} has performed an extensive study of these quasi-periodic structures in more than two dozen pulsars in all four Stokes parameters. Notably, microstructures have recently been identified in a few millisecond pulsars as well \citep[see, e.g.,][]{de2016detection,liu2022detection}, further supporting the notion that these structures are a common feature of pulsar emission. They possess a typical width, $\tau_{\mu}$, and a time period, $P_{\mu}$, with the latter falling near one thousandth of the pulsar's rotational period ($P_{rot}$). These structures have also been observed in Fast Radio Bursts \citep{refId0}, magnetars \citep{Maan2019,kramer2024quasi} as well as long period pulsars \citep{caleb2022discovery}. In addition, it has also been inferred that the micropulse width scales with $P_{rot}$ \citep[e.g., see][]{cordes1979pulsar,kramer2002high}.
Despite various studies on microstructures, their origin and connection with the pulsar emission mechanism are yet to be fully understood. Two general types of models have been considered to explain them. In the first model, often referred to as the beaming model \citep{benford1977model}, the intensity fluctuations arise from thin flux tubes that carry charged particles along the magnetic field lines emitting radiation in the direction of propagation. The angular beam width of the radiation manifests as the width of the micropulse. In the second model, called the temporal model \citep{cordes1981radio}, charged particles stream away from the pulsar in the form of radial structures which then emit radiation once they reach a certain height. Here, the width of the micropulse results from the thickness of the emitting structure. More specific models that fall into one of these broad models have also been introduced \citep[see, e.g.][]{mitra2020single,machabeli2001nature,van1980micropulses}.
Studying the microstructures, and relating them to other pulsar emission features, would provide insights into the underlying physical processes driving the emission. In the pursuit of identifying single-pulses displaying quasi-periodic narrow substructures in a pulsar time series, it requires one to inspect every individual single-pulse time series one after the other. Given that an observed pulsar time series often contains thousands to millions of single pulses, examining every single one of them manually is a rather laborious and time-intensive task. Substantially enhancing the efficiency in this process requires automated detection of microstructure candidates, robustly estimating their detection significance and screening the candidates based on their significances. Such an approach becomes even more important if microstructures were to be detected and analysed from a sizeable sample of pulsars. With this motivation, we introduce Quasi-periodic MIcrostructure Search Tool (\qmist), a Python-based software pipeline designed to robustly detect quasi-periodic microstructures from radio pulsars. In addition to demonstrating the pipeline using observations of pulsars already known to exhibit quasi-periodic microstructures, we have also conducted a survey of microstructures in 27 pulsars using \qmist. The survey has utilised \qmist on observations already available with us as well as those available from the ATNF archive, more details of which are provided later in this paper.
The rest of the paper is organised as follows. In Section \ref{sec:description}, we describe the algorithms used in \qmist along with a detailed description of the individual data processing stages in the pipeline. Section \ref{sec:survey} details the datasets used, analysis, and results of our survey of the quasi-periodic microstructures. A comprehensive discussion of the results, including the relationship between $P_{\mu}$ and $P_{\rm rot}$ derived from our findings, is presented in Section \ref{sec:discussion}. Finally, we summarise and conclude in Section \ref{sec:conclusion}.
\section{\qmist: Algorithms and detailed description}
\label{sec:description}
 
\qmist accepts a one-dimensional, dedispersed time series data in the form of a 32-bit binary file as input, like the one output from the \texttt{prepdata} utility inside \texttt{PRESTO}  \citep{ransom2011presto}. Before preparing such a time series, various preprocessing stages are important, e.g., to mitigate radio frequency interference (RFI) to improve signal-to-noise ratio (SNR) and reduce false detections. Further, it is crucial to perform dedispersion at a reasonably precise dispersion measure (DM) as a slightly incorrect DM can cause pulse smearing, making the identification of narrow periodicities challenging. Pulsar data stored in standard search-mode formats, such as the SIGPROC filterbank files, are typically processed using \texttt{PRESTO} for dedispersion and RFI mitigation. The processed frequency-scrunched one-dimensional time series data is written into a binary file (.dat) and an information file (.inf) containing relevant metadata is also generated, which \qmist utilises. In addition, the pipeline requires the precise rotational period of the pulsar at the observing epoch, along with the phase range defining the on-pulse window. These parameters can also be obtained using \texttt{PRESTO}. More details about the preprocessing stages are given in Section \ref{sec:survey}. The binary file containing the RFI-mitigated, dedispersed, intensity time series is imported to \qmist to search for single pulses exhibiting microstructures. The rotation period can either be specified explicitly, or the \texttt{bestprof} file generated by \texttt{PRESTO}'s folding process could be provided.
\subsection{Methodology}
\begin{figure}[ht!]
\centering
\includegraphics[width=0.92\linewidth]{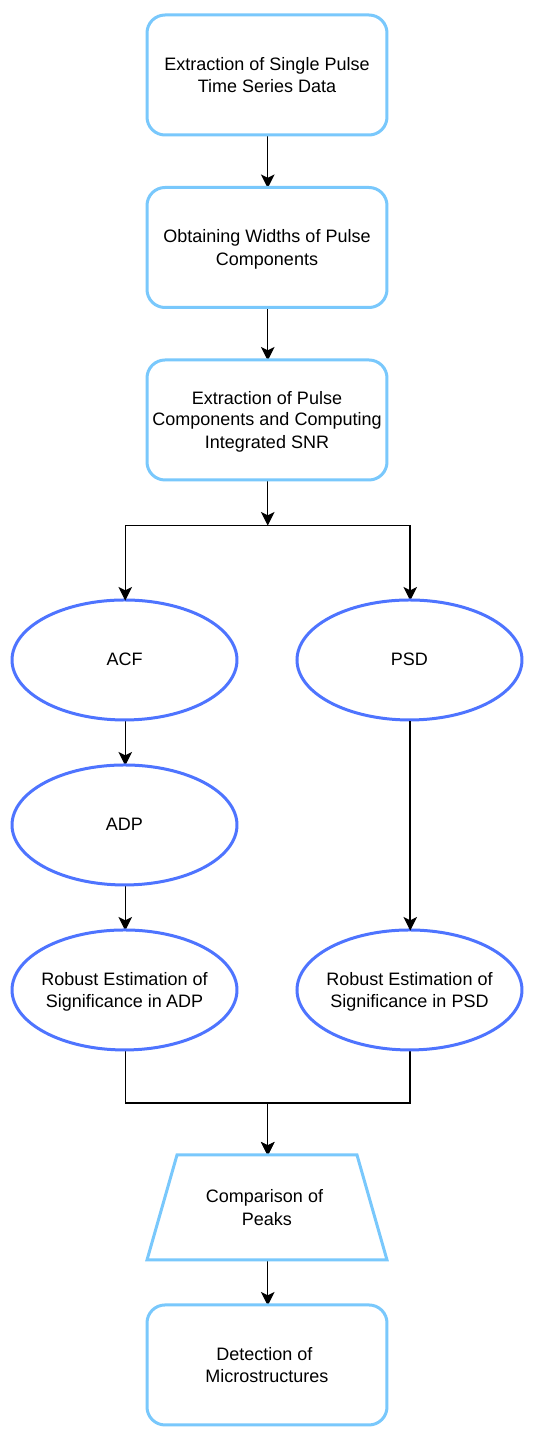}
\caption{A flowchart describing the overview of \qmist.}
\label{flowchart}
\end{figure}

In this section, we elucidate the algorithm used in the pipeline to detect microstructure periodicities in the single pulses of a pulsar emission time series. Once the data file obtained from PRESTO is imported into \qmist, it undergoes three primary processing steps. The first step involves extracting single pulses from the time series and estimating the approximate width of pulse components that make up each single pulse. Following this, the individual pulse components are isolated and their signal-to-noise ratios (SNRs) are estimated. Finally, the pipeline performs an iterative search for periodicities in each pulse component's time series that exceed a specified threshold. The following subsections provide a detailed explanation of each step. We also demonstrate the results at various processing stages of the pipeline using a time-series on pulsar B0525+21. A flowchart describing a high-level overview of the algorithm is shown in Figure \ref{flowchart}. 
\begin{figure*}[h!]
\centering
\begin{subfigure}{0.48\textwidth}
\centering
\includegraphics[width=0.84\linewidth]{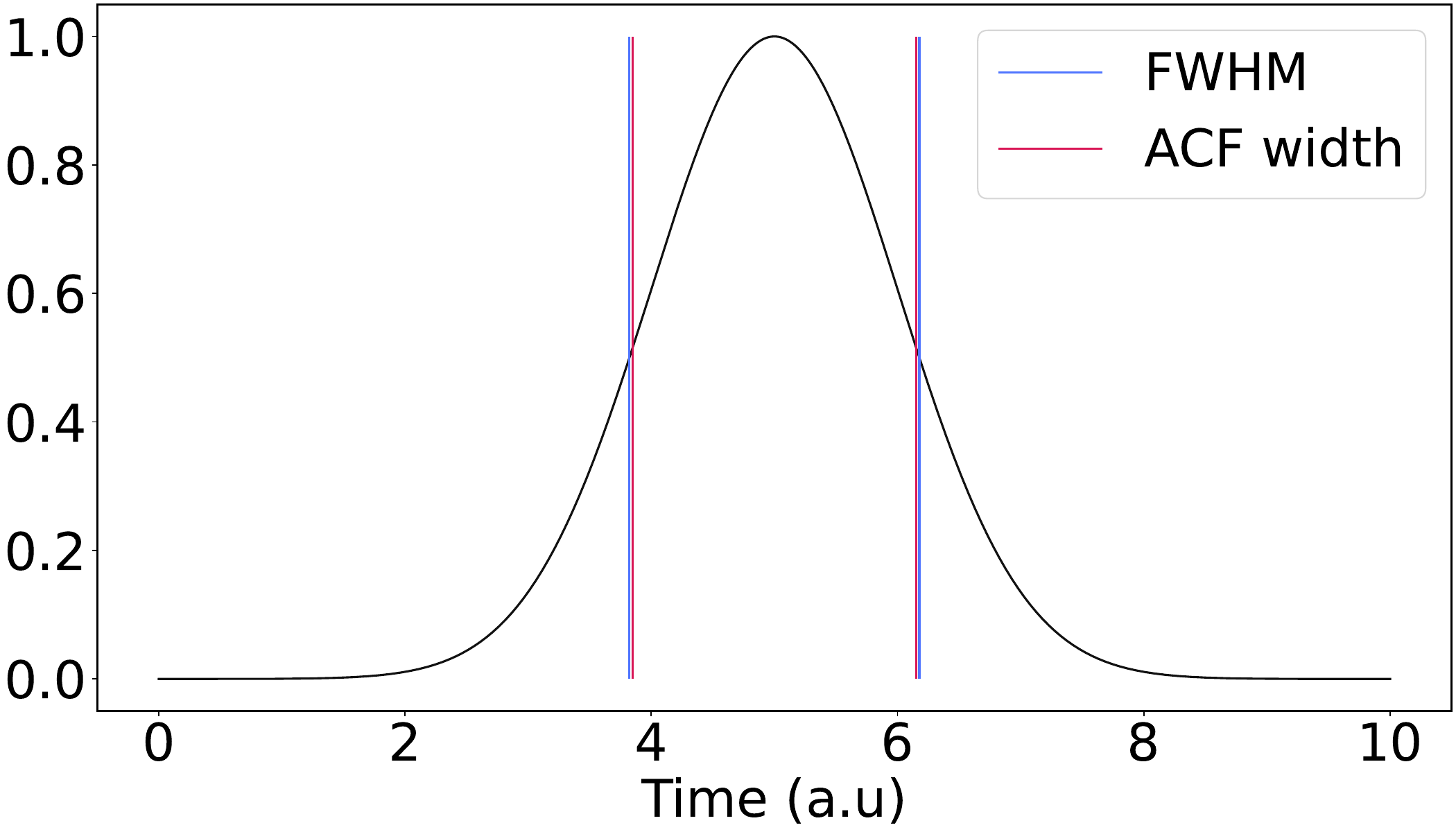}
\end{subfigure}
\begin{subfigure}{0.48\textwidth}
\centering
\includegraphics[width=1\linewidth]{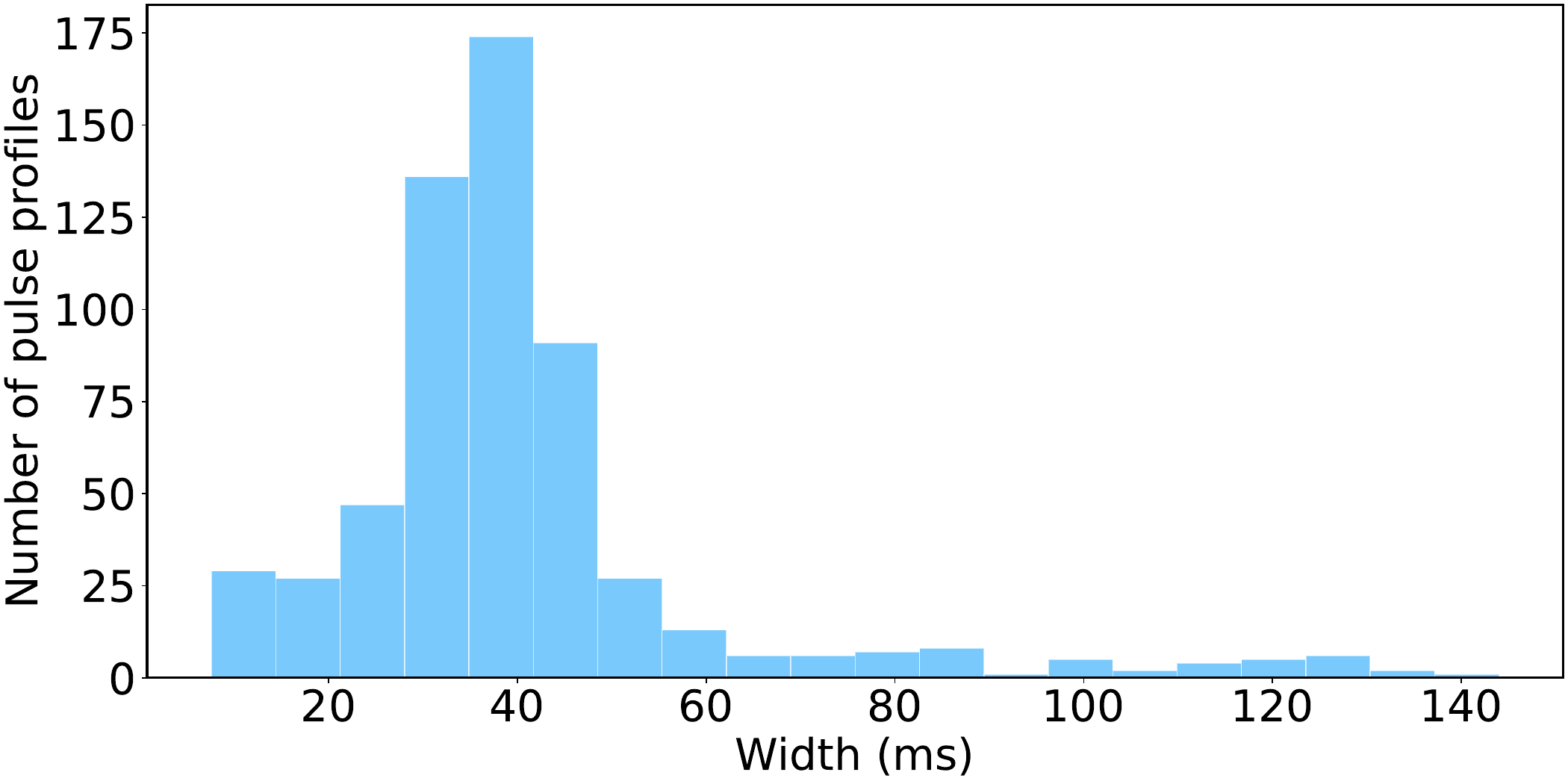}
\end{subfigure}
\caption{\textsl{Left:} A Gaussian curve with a unity standard deviation. The FWHM is indicated by the blue vertical lines, while the red vertical lines show the width obtained by the ACF analysis. \textsl{Right:} A histogram of the average pulse components widths in pulsar B0525+21.}
\label{fig-widths}
\end{figure*}
\subsubsection{Width of Pulse Components}
The initial stage of the algorithm involves extracting single-pulse profiles from the pulsar time series data within the on-pulse emission window (in units of the pulsar rotation phase) specified by the user. The user can also input the off-pulse region, in units of the rotation phase, if required. If the off-pulse region is not specified, the pipeline selects a segment outside the emission region on both sides of the specified on-pulse region, each equivalent to half the length of the on-pulse window. These two off-pulse segments are then concatenated to produce the off-pulse signal corresponding to the on-pulse signal for a given single pulse. In scenarios where the specified on-pulse region is wider than half the rotation period, the full region outside the specified on-pulse is chosen as the off-pulse region. For the processing involved in this particular stage, both the on-pulse and off-pulse signals are down-sampled by a user-specified factor (we used this factor as 4, uniformly in our analysis described later) to reduce computational load and retain only the broader features. 
After extracting the time series for all the single-pulses, the next step involves estimating the approximate width of individual pulse components within each single pulse. To achieve this, we compute the normalised autocorrelation function (ACF) of the down-sampled single-pulse time series given as follows.
\begin{equation}
\begin{aligned}
\rho(k) = \dfrac{\sum_{t=k+1}^T (y_{t} - \bar{y})(y_{t-k} - \bar{y})}{\sum_{t=1}^T (y_t - \bar{y})^2}
\end{aligned}
\end{equation}
where $y_t$ is the intensity at $t^{th}$ bin, $\bar{y}$ is the mean of the time series, $k$ is the lag which starts at 0, and T is the total sample size of the time series. Note that $\rho(0)$=1 by design, and it is often dominated by contamination from noise. So, we use $\rho(1)$ as a proxy for the maximum $\rho$. We determine the value of $k$ at which the ACF reaches half the value at $k=1$. This lag is considered to be approximately equivalent to half the average width of the pulse components within the single pulse, and is subsequently doubled to obtain the full width. Figure~\ref{fig-widths} shows a comparison between the width obtained from the above ACF analysis and the actual full-width-half-maximum (FWHM) of a Gaussian curve. Figure~\ref{fig-widths} also shows the distribution, as an example, of the resultant widths from an observation of the pulsar B0525+21. 
\subsubsection{Signal-to-Noise Ratio estimation}
\begin{figure*}[h!]
\centering
\begin{subfigure}{0.75\linewidth}
\centering
\includegraphics[width=1\linewidth]{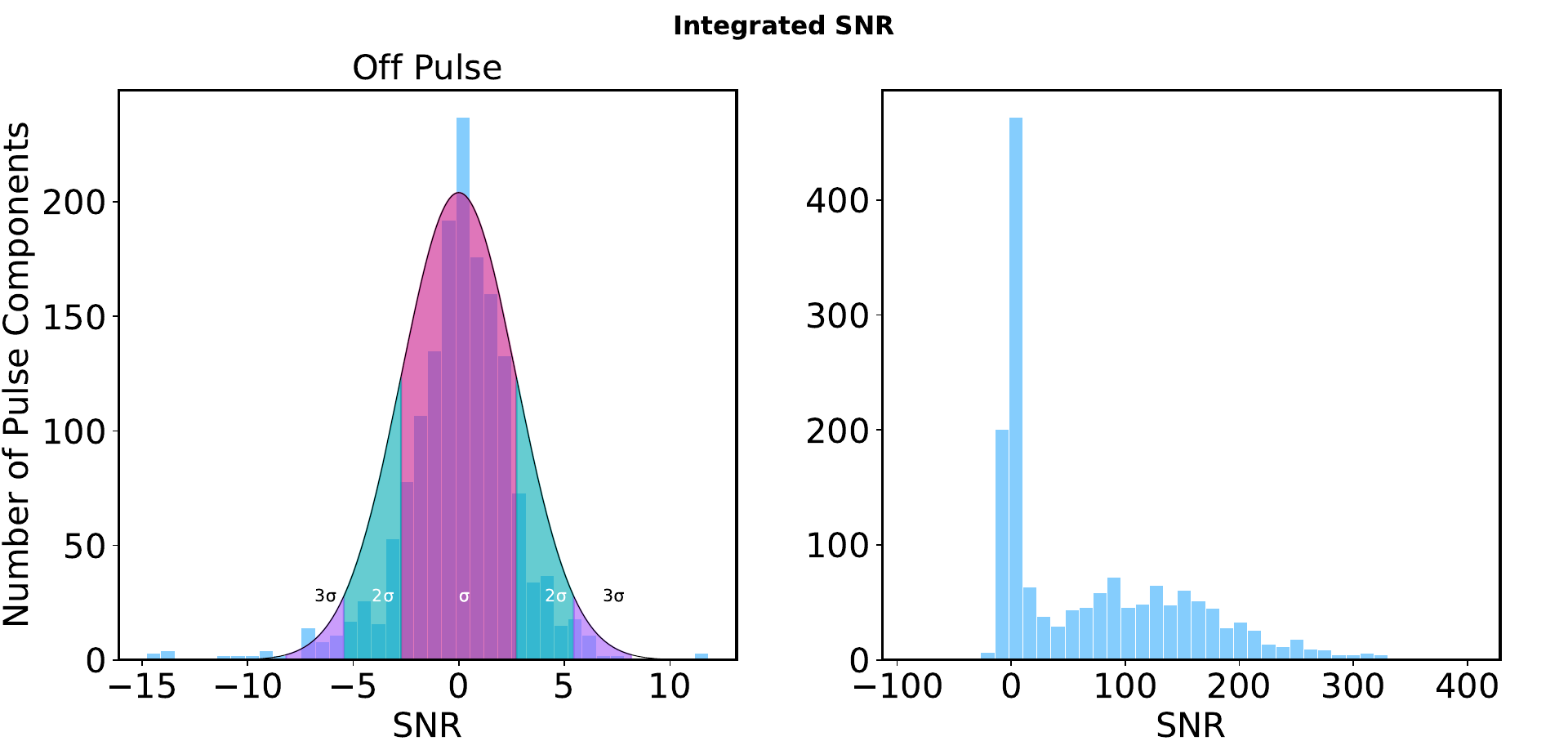}
\end{subfigure}
\begin{subfigure}{0.75\linewidth}
\centering
\includegraphics[width=1\linewidth]{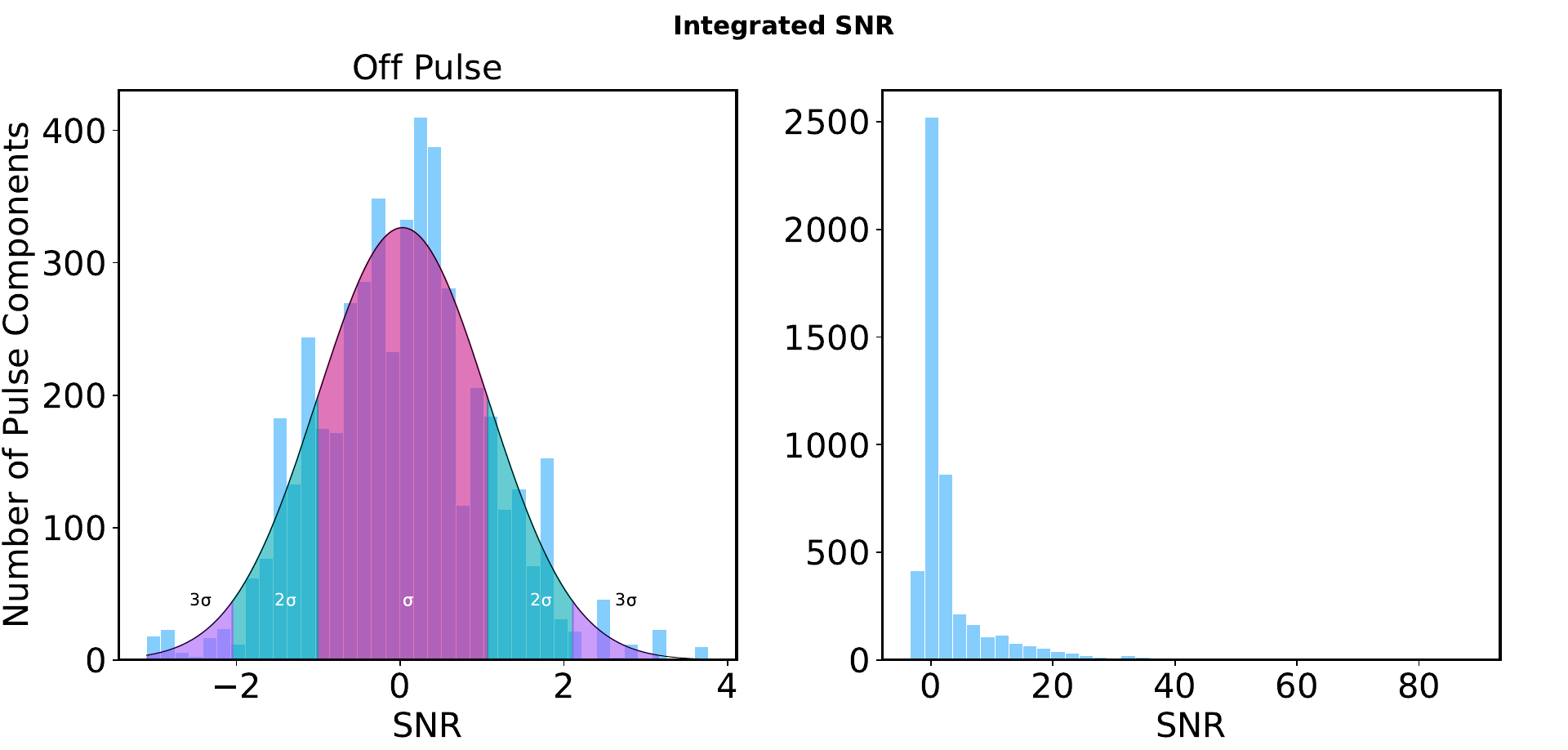}
\end{subfigure}
\caption{The SNR distributions of the off-pulse (left) and on-pulse (right) signals of pulsars B0525+21 (top) and B0540+23 (bottom). A Gaussian curve fitted to the integrated SNR histograms of the off-pulse regions are also shown, with the regions separated by unit standard deviations coloured differently. The SNR distributions for B0525+21 are heavily contaminated by baseline variations.}
\label{snrhist}
\end{figure*}

In the second stage, the individual pulse components within a single-pulse profile are extracted. The individual single-pulse time series is cross-correlated with a boxcar function of width equal to the respective width obtained in the previous stage. Then, the locations of the pulse components are determined from the peaks in the cross-correlation function. Once all the pulse component time series are extracted from the whole observation, they are converted to units of SNR using the following.
\begin{equation}
\begin{aligned}
SNR(i) = \dfrac{p_i - \bar{p}_{off}}{\sigma_{off}}
\end{aligned}
\end{equation}
where $p$ is the intensity of the $i^{th}$ bin, $\bar{p}_{off}$ and $\sigma$ are the median and the RMS of the off-pulse signal. Following \citet{lorimer2005handbook}, the integrated SNR is estimated as follows.
\begin{equation}
\begin{aligned}
SNR_{int} = \dfrac{\sum p_i - \bar{p}_{off}}{\sigma_{off}\sqrt{w}}
\end{aligned}
\end{equation}
where $w$ is the pulse width estimated earlier, in units of the time samples.
The locations and the integrated SNRs of the extracted pulse components are stored into a file for the next processing stage. The SNR$_{int}$ of both the off-pulse and the on-pulse regions are calculated and their distributions are examined. While the SNR$_{int}$ of the off-pulse regions should be close to zero and should follow random distribution with a standard deviation of 1, these could vary significantly from pulse to pulse due to, e.g., low-level RFI and baseline variations at timescales close the pulse-width. Examples of such histograms are shown in Figure \ref{snrhist} for the pulsars B0525+21 and B0540+23. As evident from the B0525+21 histograms, the baseline variations could cause underestimation as well as overestimation of SNR, and thus potentially missing genuinely high-SNR pulses and picking up low-SNR pulses. Thus, it would be important to correct for such variations in advance before searching for quasi-periodic microstructures. 
\subsubsection{Detection of microstructure periodicities and robust significance estimates}
Now, we have reached the final processing stage of our pipeline, where it is able to fulfil its intended purpose. The algorithm for detecting periodicities in a given pulse component time series largely adheres to the methodology outlined in \citet{lange1998radio}. First, for each of the pulse components that has integrated SNR larger than a specified threshold, the original time series is used to extract a portion centred at the pulse location and as long as twice the width of the component. Then, for each of these pulse components, the ACF and its derivative are computed utilising the central 75\% of the extracted portion. Since the first point of the ACF, i.e., corresponding to zero lag, is equal to 1.0 by design, and the next point would often show a sharp drop,  we replace the zero lag point value by that at the first lag to avoid this sharp discontinuity in ACF and the corresponding large change in its derivative. This adjustment also prevents the discontinuity from introducing unwanted artefacts in subsequent steps. Extrema in the ACF which might correspond to periodicities are identified by changes in the sign of the slope (i.e., the turn-over points) in the ACF derivative or using a Fourier transform. However, the ACF and its derivative generally also exhibit an underlying baseline variation or red noise due to the broad nature of the pulse components. To get rid of any underlying broad variations, the ACF derivative is fitted with a fourth-order polynomial, which is then subtracted to obtain a linearised ACF derivative. Subsequently, the ACF derivative power spectrum (ADP) is obtained using the Fourier transform:
\begin{equation}
\begin{aligned}
ADP = \left|FT\left (\dfrac{d}{dt} \rho(t)\right )\right|^2
\end{aligned}
\end{equation}
where FT denotes the Fourier transform, which is computed using the Fast Fourier Transform (FFT).
Independently, slow baseline variations from the pulse component time series are also removed by subtracting a best-fit fourth-order polynomial, and then the power spectral density (PSD) is derived by performing a Fourier transform on the residual. Any periodicity present in the time series would produce a peak in the PSD as well as the ADP. We note that a slightly different approach was introduced by \citet{mitra2015polarized} to separate the narrow quasi-periodic peaks from the broad sub-pulse. Their method involved a spline fit on the pulse component instead of the fourth-degree polynomial used in our method. \citet{singh2024single} employed a running median filter for the same step.
\begin{figure}[ht!]
\centering
\includegraphics[width=1\linewidth]{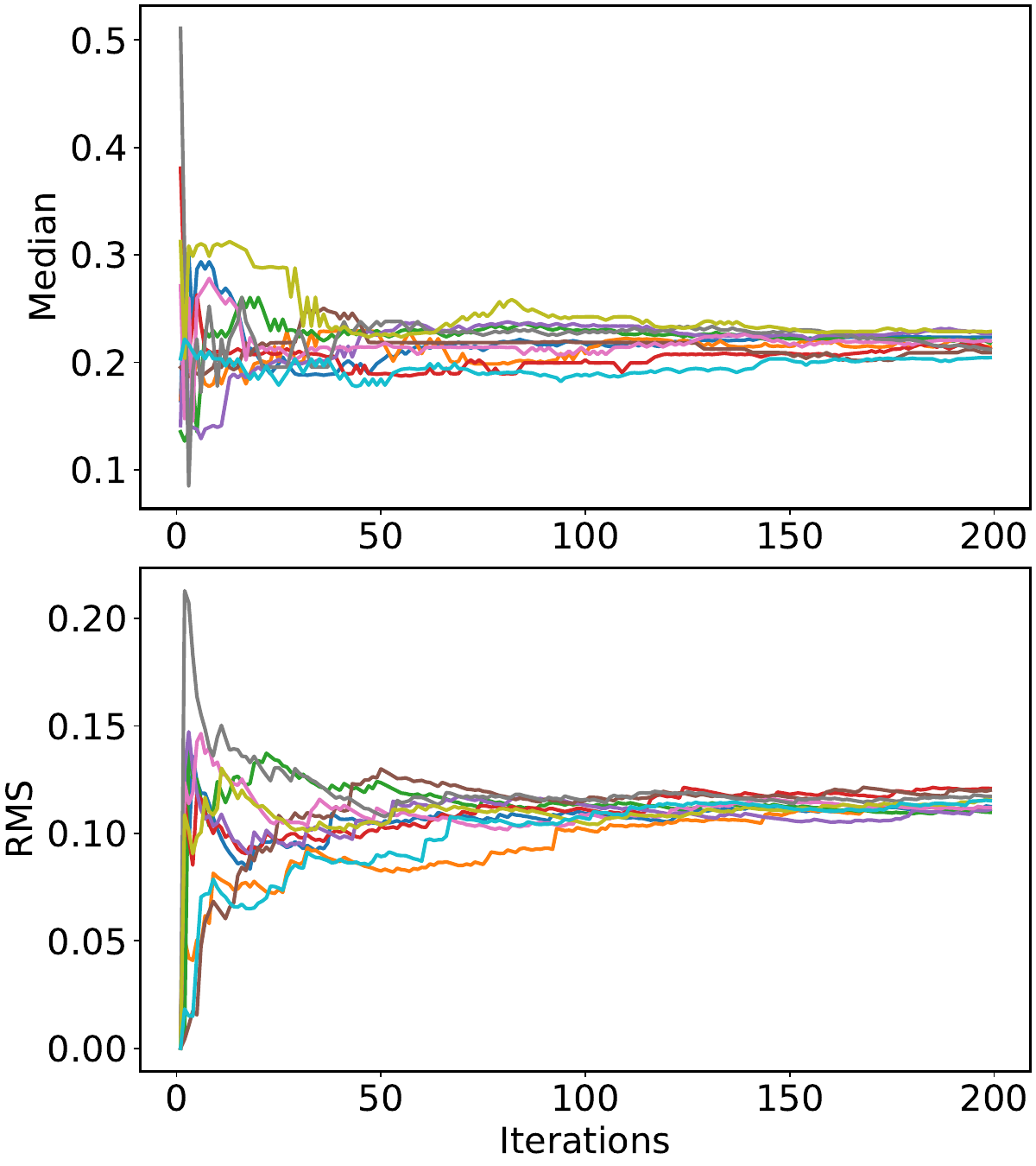}
\caption{Variations of the median and RMS at the first 10 frequency bins in the ADP are shown in different colours and as a function of iterations.}
\label{medrms}
\end{figure}
\begin{figure*}[ht!]
\centering
\includegraphics[width=1\linewidth]{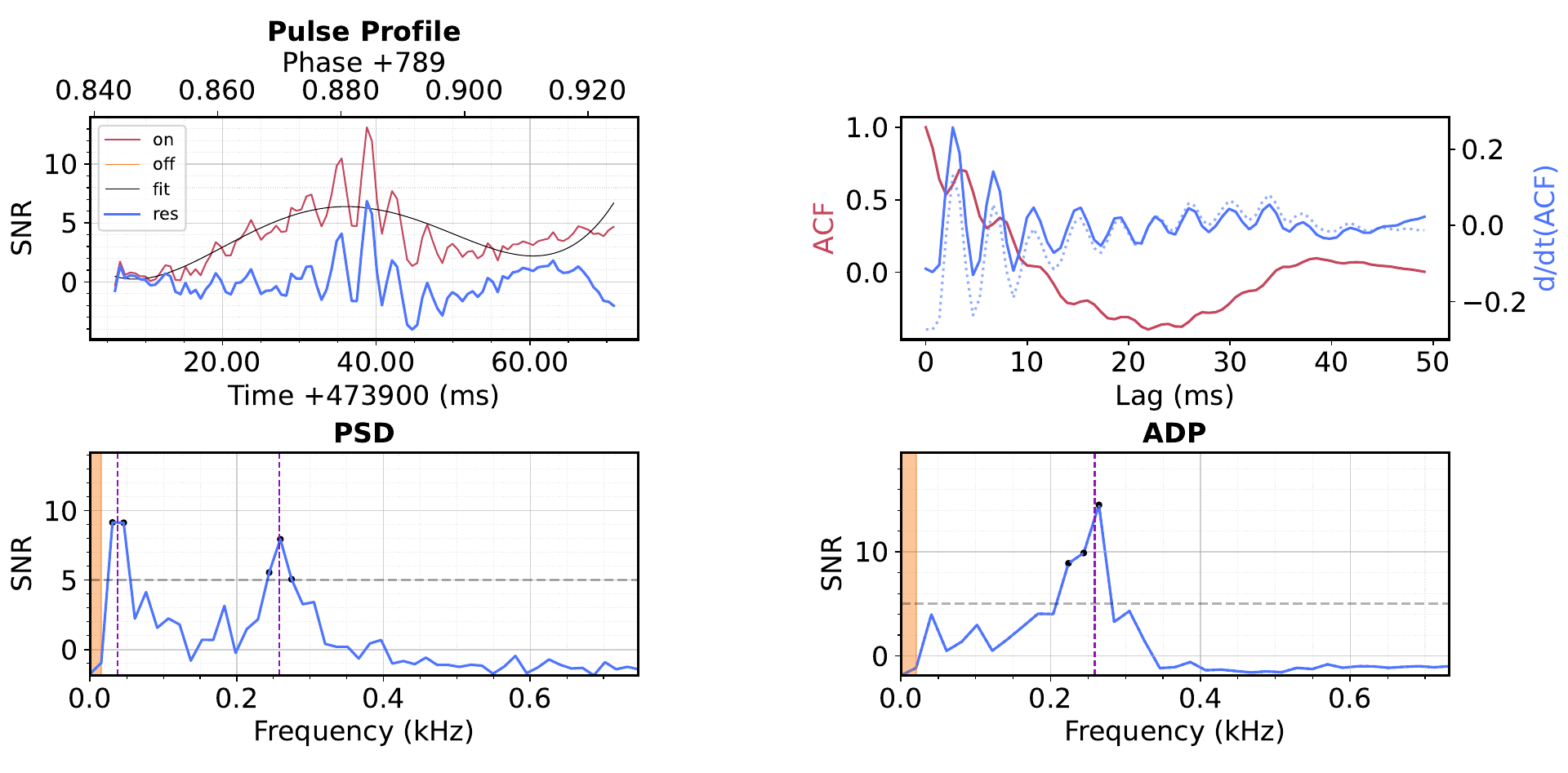}
\caption{An example diagnostic plot produced by \qmist for a single pulse component observed from pulsar B0525+21. The top-left panel shows the on-pulse (brown) and off-pulse (orange) time series along with a 4$^{th}$-degree polynomial (black) fitted to the pulse component, and the residual time series (blue). The top-right panel shows the ACF (red) and ACF-derivative (blue), along with the ACF-derivative after fitting and subtracting a 4$^{th}$-degree polynomial (dotted blue). The bottom-left and bottom-right panels display the PSD and ADP, respectively, in units of SNR. The purple vertical lines mark the peaks that are above the specified threshold (black dashed line). The orange shade in both these plots covers the frequencies which are masked.}
\label{result}
\end{figure*}
To identify genuine periodicities in the pulse time series or its ACF and ACF-derivative, it is crucial to robustly determine the detection significance of every point in the ADP and PSD. As both of these quantities represent powers in the Fourier domain, the statistics could differ as a function of Fourier frequency. To robustly estimate the detection significance of the points in the ADP, we utilise a bootstrap method. First, the first few data points (which would correspond to the overall width of the pulse component) are zeroed out in both the real and imaginary parts of the ADP separately. Then the data points lying above $2.5\sigma$ are zeroed out as well. Next, the inverse Fourier transform is taken to produce a new ACF derivative. We then shuffle the data points of this new ACF randomly. Two hundred such randomly shuffled ACF samples are generated and their corresponding ADPs are computed. Utilising the sample of these 200 ADPs, the median, $F_{mean}$, and RMS, $F_{RMS}$, are determined separately for each Fourier frequency bin. The convergence of the bin-wise mean and RMS against the number of randomly shuffled ACF samples is illustrated in Figure \ref{medrms}. Based on this, we considered two hundred to be an appropriate number of random samples. Using the above estimated Fourier frequency-dependent median and RMS, we assign an SNR, $ADP_{SNR}$, to each point in the original ADP as follows. 
\begin{equation}
\begin{aligned}
ADP_{SNR} = \dfrac{ADP - F_{mean}}{F_{RMS}}
\end{aligned}
\end{equation}
The same procedure is performed on the PSD as well, to derive the quantity $PSD_{SNR}$. $ADP_{SNR}$ and $PSD_{SNR}$ are robust estimates of how strong any periodicity feature in the corresponding Fourier domains would be compared to the respective noise levels. We also note that the ADP and PSD in units of $ADP_{SNR}$ and $PSD_{SNR}$ could also look slightly different from the original ADP and PSD, respectively, depending on how the median and RMS changes as a function of the Fourier frequency. At times, presenting the ADP and PSD in units of these robust SNRs could also bring out significant peaks otherwise obscured within a dense array of noise-related peaks.
Before identifying significant peaks in ADP and PSD, the first few data points corresponding to the overall width of the pulse component are masked out as these points could still have contributions from the slow variations over the pulse width. This process was carried out by considering the width obtained in the previous stage as the approximate FWHM of the pulse component, assuming a Bell shape, and deriving its counterpart in the Fourier space. The FWHM $w$ of a bell-shaped function in terms of its standard deviation $\sigma$ is given by,
\begin{equation}
\begin{aligned}
w = 2\sqrt{2ln(2)}\sigma
\end{aligned}
\label{width}
\end{equation}
The relation between the standard deviation in the time domain and that in the frequency domain is given by (see Appendix \ref{sec:sigsig}),
\begin{equation}
\begin{aligned}
\sigma_t \sigma_f = \dfrac{1}{2\pi}
\end{aligned}
\label{sigtsigf}
\end{equation}
where the subscripts $t$ and $f$ denote the time and frequency domain, respectively.
Solving for $w_f$ in terms of $w_t$ using equations \ref{width} and \ref{sigtsigf}, we get
\begin{equation}
\begin{aligned}
    w_f = \dfrac{4ln(2)}{w_t\pi}
\end{aligned}
\end{equation}
The bins up to this width are masked out in the Fourier domains. This ensures that only the peaks due to genuine periodicities are considered. 
Next, the locations of the peaks exceeding a specified threshold value are selected using \texttt{find\_peaks} function from the Scipy library \citep{2020SciPy-NMeth} with the height parameter set to the desired threshold value and prominence parameter set to 1. To estimate the Fourier frequency, and thus the associated period value, one neighbouring point on either side of the identified maximum is selected, and sinc interpolation is applied on these three points. The Fourier frequency at the maximum of the fitted sinc curve is then assigned as the peak frequency. A fraction of the peak features sometimes exhibit a small dip near the peak. To obtain the peak frequency for such features, a weighted average of the points near the identified peak is computed. Subsequently, we compare various peaks identified independently in $ADP_{SNR}$ and $PSD_{SNR}$, following a similar scheme used by \citet{kramer2024quasi}. More specifically, if the corresponding peak frequencies, $f_{PSD}$ and $f_{ADP}$, differ by no more than 20\%, i.e., if $\dfrac{|f_{PSD} - f_{ADP}|}{0.5\times(f_{PSD} + f_{ADP})} < 0.2$, we assume that the peak frequency corresponds to a genuine feature corresponding to periodic structures in the pulse component.
The whole process is carried out sequentially on all the pulse components. For the components found to exhibit microstructures, the plots of their time series, ACF and ACF-derivative, PSD, and ADP are saved as diagnostics into a PDF file, and their microstructure periods, SNRs and other information are saved into a CSV file. An example diagnostic plot is shown in Figure~\ref{result}.
\begin{figure}[hb!]
\centering
\includegraphics[width=1\linewidth]{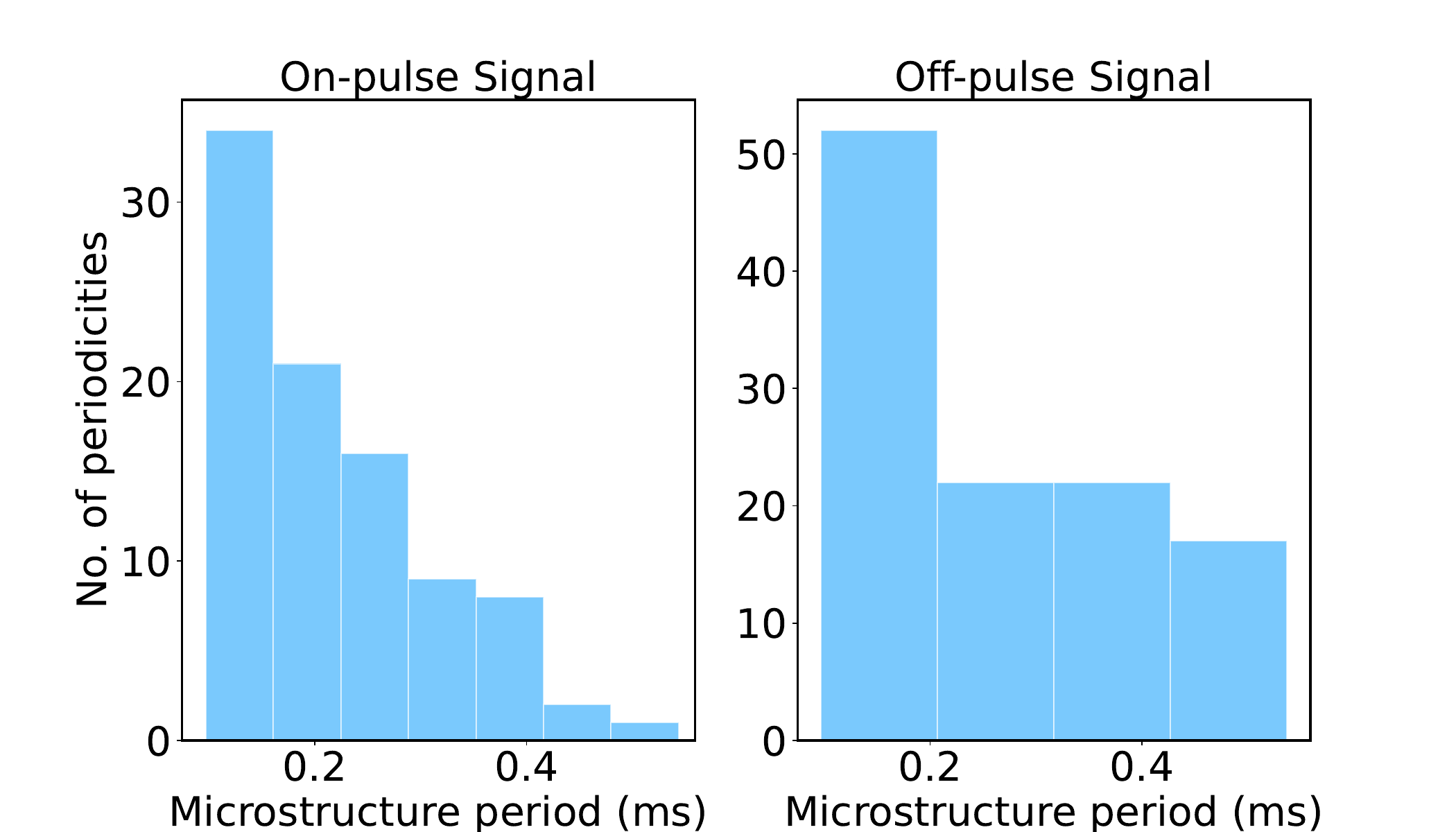}
\caption{Distributions of the quasi-periods found by \qmist from the on-pulse (left) and off-pulse (right) regions in the time series of B1706$-$44.}
\label{b1706hist}
\end{figure}

\subsubsection{Periodicities Due to RFI}
While \qmist will pick up microstructure periodicities in a pulse component, it would also detect any periodic artefacts that could have a different origin if the periodicity is strong enough. These artefacts are generally caused by RFI, in addition to any statistical outliers, and such cases would add to the false positives. RFI-originated artefacts could be caused by an interference that could be external to the system or from components in the system (e.g., leakage of 50\,Hz or other clock signals). However, such artefacts, whether caused by RFI or statistical outliers, would be detectable in the off-pulse as well as on-pulse regions. Figure~\ref{b1706hist} shows one such example where the distributions of the periodicities detected from the on-pulse as well as off-pulse regions span a similar range of periods. Therefore, while \qmist significantly reduces the number of pulses that need to be examined, a manual examination of the diagnostic plots of the candidate pulses output from the pipeline is essential to distinguish genuine microstructures from any artefacts in the data.
\section{A survey of quasi-periodic microstructures in pulsars}\label{sec:survey}
We have used \qmist to conduct a survey of quasi-periodic microstructures in pulsars, using our own existing observations as well as the archival data. In the following subsections, we provide the details of our sample selection, data collection and results obtained from this survey.
\subsection{Sample selection, observations and processing}
The pulsar observations we have utilised in our survey were conducted using the Giant Metrewave Radio Telescope (GMRT), the Green Bank Telescope (GBT), and the Parkes telescope. The observations of 15 pulsars using the GMRT and GBT that we have utilised here were originally conducted with different scientific objectives. Some of these pulsars were observed as test pulsars or control observations, and hence resulted in relatively small scans. Majority of these test pulsars come from two observing programs, a long-term follow-up of the magnetar XTE~J1810$-$197 \citep{Maan2019,Maan2022} and the SCOPE pulsar survey \citep{Maan2026a}. In addition, we have utilised the archival observations from the Murriyang, CSIRO's Parkes telescope. First we utilised the ATNF pulsar catalogue \citep{atnf} to find pulsars that are relatively bright at 1.4\,GHz (flux density more than 1\,mJy), constitute the part of the young pulsar population (rotation periods between 50\,ms and 300\,ms), and have DM less than 100\,pc\,cm$^{-3}$. The selection criteria on rotation periods and DM are motivated to ensure that the dispersion smearing (across a channel) is less than one-thousandth of a period as well as to minimise the smearing due to interstellar scattering. Our search resulted in 34 pulsars and a search on the CSIRO Data Access Portal resulted in availability of useful data for 13 of these pulsars, including one (PSR~B1706$-$44) for which GMRT data were already available. Whenever possible, data from more than one observation were downloaded and utilised. The majority of the previous microstructure studies focused on pulsars located in the Northern Hemisphere, and utilising the data from the Parkes telescope in our survey broadens the scope to Southern pulsars.
\begin{sidewaystable*}[!htp]
\renewcommand{\arraystretch}{0.8}
\setlength{\tabcolsep}{4pt}
\caption{A summary of targets and observations used in our survey and the results obtained.}
\label{tab:survey}
\begin{tabular}{ccccccccccccccc}
\midrule
\cellcolor{headerblue}\textcolor{white}{\textbf{Source}} &\cellcolor{headerblue}\textcolor{white}{\textbf{P (s)}} &\cellcolor{headerblue}\textcolor{white}{\textbf{\makecell{DM \\(pc/cc)}}} &\cellcolor{headerblue}\textcolor{white}{\textbf{Telescope}}
&\cellcolor{headerblue}\textcolor{white}{\textbf{\makecell{Epoch \\(MJD)}}} &\cellcolor{headerblue}\textcolor{white}{\textbf{\makecell{CF \\(MHz)}}}
&\cellcolor{headerblue}\textcolor{white}{\textbf{\makecell{$T_{samp}$ \\($\mu$s)}}}
&\cellcolor{headerblue}\textcolor{white}{\textbf{\makecell{Obs. \\ length \\(periods)}}} &\cellcolor{headerblue}\textcolor{white}{\textbf{\makecell{$n_{pulses}$ \\ SNR$>$10}}} &\cellcolor{headerblue}\textcolor{white}{\textbf{\makecell{Pulses \\ with MS \\ (actual/output)}}} &\cellcolor{headerblue}\textcolor{white}{\textbf{\makecell{Pulses \\ with MS \\ (\%)}}} &\cellcolor{headerblue}\textcolor{white}{\textbf{$P_{\mu}$ ($\mu$s)}} &\cellcolor{headerblue}\textcolor{white}{\textbf{\makecell{$P_{\mu}$ from \\ literature ($\mu$s)}}} &\cellcolor{headerblue}\textcolor{white}{\textbf{Reference}}\\
J0134-2937 &0.137 &21.8 &Parkes &57140 &1382 &64 &2803 &183 &0/2 &0 &-- &-- &-- \\
& & &Parkes &57245 &1382 &64 &1960 &132 &0/3 &0 &-- & & \\
\cellcolor{lightblue}J0407+1607 &\cellcolor{lightblue}0.026 &\cellcolor{lightblue}35.6 &\cellcolor{lightblue}GBT &\cellcolor{lightblue}60371 &\cellcolor{lightblue}820 
&\cellcolor{lightblue}82 &\cellcolor{lightblue}11779 &\cellcolor{lightblue}0 &\cellcolor{lightblue}-- &\cellcolor{lightblue}-- &\cellcolor{lightblue}-- &\cellcolor{lightblue}-- &\cellcolor{lightblue}-- \\
B0525+21 &3.746 &50.9 &GMRT &58787 &650 &655 &641 &508 &64/443 &12.6 &8633 $\pm$ 1370 &5650 $\pm$ 3880 &(M) \\
& & &GMRT &58787 &650 &655 &646 &559 &73/506 &13.1 &9251 $\pm$ 1596 & & \\
\cellcolor{lightblue}B0540+23 &\cellcolor{lightblue}0.246 &\cellcolor{lightblue}77.7 &\cellcolor{lightblue}Parkes &\cellcolor{lightblue}57129 &\cellcolor{lightblue}1382
&\cellcolor{lightblue}64 &\cellcolor{lightblue}2014 &\cellcolor{lightblue}1488 &\cellcolor{lightblue}83/609 &\cellcolor{lightblue}5.6 &\cellcolor{lightblue}821 $\pm$ 144 &\cellcolor{lightblue}-- &\cellcolor{lightblue}(L) \\
\cellcolor{lightblue} &\cellcolor{lightblue} &\cellcolor{lightblue} &\cellcolor{lightblue}Parkes &\cellcolor{lightblue}57244 &\cellcolor{lightblue}1382 
&\cellcolor{lightblue}64 &\cellcolor{lightblue}786 &\cellcolor{lightblue}459 &\cellcolor{lightblue}38/112 &\cellcolor{lightblue}8.3 &\cellcolor{lightblue}790 $\pm$ 225 &\cellcolor{lightblue} &\cellcolor{lightblue} \\
\cellcolor{lightblue} &\cellcolor{lightblue} &\cellcolor{lightblue} &\cellcolor{lightblue}Parkes &\cellcolor{lightblue}57101 &\cellcolor{lightblue}1382 
&\cellcolor{lightblue}64 &\cellcolor{lightblue}875 &\cellcolor{lightblue}789 &\cellcolor{lightblue}6/120 &\cellcolor{lightblue}0.8 &\cellcolor{lightblue}733 $\pm$ 107 &\cellcolor{lightblue} &\cellcolor{lightblue} \\
\cellcolor{lightblue} &\cellcolor{lightblue} &\cellcolor{lightblue} &\cellcolor{lightblue}Parkes &\cellcolor{lightblue}57935 &\cellcolor{lightblue}1382
&\cellcolor{lightblue}64 &\cellcolor{lightblue}786 &\cellcolor{lightblue}384 &\cellcolor{lightblue}25/66 &\cellcolor{lightblue}6.5 &\cellcolor{lightblue}990 $\pm$ 361 &\cellcolor{lightblue} &\cellcolor{lightblue} \\
J0631+1036 &0.288 &125.3 &GBT &60342 &820 &82 &2103 &58 &0/6 &0 &-- &-- &-- \\
\cellcolor{lightblue}B0727-18 &\cellcolor{lightblue}0.510 &\cellcolor{lightblue}61.3 &\cellcolor{lightblue}GMRT &\cellcolor{lightblue}57979 &\cellcolor{lightblue}650 
&\cellcolor{lightblue}82 &\cellcolor{lightblue}359 &\cellcolor{lightblue}51 &\cellcolor{lightblue}0/4 &\cellcolor{lightblue}0 &\cellcolor{lightblue}-- &\cellcolor{lightblue}-- &\cellcolor{lightblue}-- \\
B0833-45 &0.089 &67.7 &Parkes &57227 &1382 &64 &2165 &2090 &103/1883 &4.9 &592 $\pm$ 133 &470 $\pm$ 160 &(K) \\
\cellcolor{lightblue}B0919+06 &\cellcolor{lightblue}0.431 &\cellcolor{lightblue}27.3 &\cellcolor{lightblue}GBT &\cellcolor{lightblue}60342 &\cellcolor{lightblue}820 
&\cellcolor{lightblue}82 &\cellcolor{lightblue}703 &\cellcolor{lightblue}696 &\cellcolor{lightblue}215/627 &\cellcolor{lightblue}30.9 &\cellcolor{lightblue}1066 $\pm$ 148 &\cellcolor{lightblue}960 $\pm$ 270 &\cellcolor{lightblue}(M) \\
B0950+08 &0.253 &2.7 &Parkes &57044 &1382 &64 &2225 &1663 &113/1004 &6.8 &798 $\pm$ 123 &630 $\pm$ 220 &(M) \\
& & &Parkes &57452 &1369 &64 &3560 &2566 &30/360 &0.8 &2412 $\pm$ 421 & 500 $\pm$ 148 & (Li) \\
& & &Parkes &60442 &1216 &64 &84 &51 &15/33 &29.4 &536 $\pm$ 132 & & \\
\cellcolor{lightblue}B1039-19 &\cellcolor{lightblue}1.386 &\cellcolor{lightblue}33.8 &\cellcolor{lightblue}GBT &\cellcolor{lightblue}60342 &\cellcolor{lightblue}820 
&\cellcolor{lightblue}82 &\cellcolor{lightblue}66 &\cellcolor{lightblue}46 &\cellcolor{lightblue}0/1 &\cellcolor{lightblue}0 &\cellcolor{lightblue}-- &\cellcolor{lightblue}-- &\cellcolor{lightblue}-- \\
J1045-4509 &0.007 &58.1 &GMRT &60556 &650 &5 &33755 &296 &0/29 &0 & &-- &-- \\
\cellcolor{lightblue}B1055-52 &\cellcolor{lightblue}0.197 &\cellcolor{lightblue}29.9 &\cellcolor{lightblue}Parkes &\cellcolor{lightblue}56123 &\cellcolor{lightblue}732
&\cellcolor{lightblue}256 &\cellcolor{lightblue}6171 &\cellcolor{lightblue}662 &\cellcolor{lightblue}0/30 &\cellcolor{lightblue}0 &\cellcolor{lightblue}-- &\cellcolor{lightblue}-- &\cellcolor{lightblue}-- \\
B1221-63 &0.217 &87.1 &Parkes &56123 &3096 &256 &5619 &247 &0/8 &0 &-- &-- &-- \\
& & &Parkes &56309 &732 &256 &5619 &363 &0/67 &0 &-- & & \\
& & &Parkes &56123 &732 &256 &5619 &340 &0/23 &0 &-- & & \\
\cellcolor{lightblue}B1317-53 &\cellcolor{lightblue}0.280 &\cellcolor{lightblue}97.1 &\cellcolor{lightblue}Parkes &\cellcolor{lightblue}57228 &\cellcolor{lightblue}1382 
&\cellcolor{lightblue}64 &\cellcolor{lightblue}691 &\cellcolor{lightblue}2 &\cellcolor{lightblue}0/0 &\cellcolor{lightblue}0 &\cellcolor{lightblue}-- &\cellcolor{lightblue}-- &\cellcolor{lightblue}-- \\
\cellcolor{lightblue} &\cellcolor{lightblue} &\cellcolor{lightblue} &\cellcolor{lightblue}Parkes &\cellcolor{lightblue}57165 &\cellcolor{lightblue}1382
&\cellcolor{lightblue}64 &\cellcolor{lightblue}691 &\cellcolor{lightblue}1 &\cellcolor{lightblue}0/0 &\cellcolor{lightblue}0 &\cellcolor{lightblue}-- &\cellcolor{lightblue} &\cellcolor{lightblue} \\
\cellcolor{lightblue} &\cellcolor{lightblue} &\cellcolor{lightblue} &\cellcolor{lightblue}Parkes &\cellcolor{lightblue}57909 &\cellcolor{lightblue}1382
&\cellcolor{lightblue}64 &\cellcolor{lightblue}691 &\cellcolor{lightblue}14 &\cellcolor{lightblue}0/0 &\cellcolor{lightblue}0 &\cellcolor{lightblue}-- &\cellcolor{lightblue} &\cellcolor{lightblue} \\
B1451-68 &0.263 &8.6 &Parkes &56866 &1382 &64 &1881 &982 &2/98 &0.2 &-- &-- &-- \\
& & &Parkes &57129 &1382 &64 &1881 &461 &8/176 &1.7 &761 $\pm$ 116 & & \\
& & &Parkes &56889 &1382 &64 &1866 &1692 &0/443 &0 &-- &-- &-- \\
\cellcolor{lightblue}B1541-52 &\cellcolor{lightblue}0.179 &\cellcolor{lightblue}35.2 &\cellcolor{lightblue}Parkes &\cellcolor{lightblue}57228 &\cellcolor{lightblue}1382
&\cellcolor{lightblue}64 &\cellcolor{lightblue}1083 &\cellcolor{lightblue}234 &\cellcolor{lightblue}0/18 &\cellcolor{lightblue}0 &\cellcolor{lightblue}-- &\cellcolor{lightblue}-- &\cellcolor{lightblue}-- \\
J1549-4848 &0.288 &56.0 &Parkes &56866 &1382 &64 &681 &1 &0 &0 &-- &-- &-- \\
& & &Parkes &57992 &1382 &64 &681 &0 &-- &-- &-- & & \\
\cellcolor{lightblue}B1706-16 &\cellcolor{lightblue}0.653 &\cellcolor{lightblue}24.9 &\cellcolor{lightblue}GMRT &\cellcolor{lightblue}60636 &\cellcolor{lightblue}1360
&\cellcolor{lightblue}40 &\cellcolor{lightblue}280 &\cellcolor{lightblue}213 &\cellcolor{lightblue}4/30 &\cellcolor{lightblue}1.9 &\cellcolor{lightblue}923 $\pm$ 98 &\cellcolor{lightblue}-- &\cellcolor{lightblue}-- \\
\cellcolor{lightblue} &\cellcolor{lightblue} &\cellcolor{lightblue} &\cellcolor{lightblue}Parkes &\cellcolor{lightblue}58012 &\cellcolor{lightblue}1382 
&\cellcolor{lightblue}64 &\cellcolor{lightblue}296 &\cellcolor{lightblue}258 &\cellcolor{lightblue}8/80 &\cellcolor{lightblue}3.1 &\cellcolor{lightblue}808 $\pm$ 214 &\cellcolor{lightblue} &\cellcolor{lightblue} \\
\cellcolor{lightblue} &\cellcolor{lightblue} &\cellcolor{lightblue} &\cellcolor{lightblue}Parkes &\cellcolor{lightblue}57309 &\cellcolor{lightblue}1382
&\cellcolor{lightblue}64 &\cellcolor{lightblue}296 &\cellcolor{lightblue}61 &\cellcolor{lightblue}0/1 &\cellcolor{lightblue}0 &\cellcolor{lightblue}-- &\cellcolor{lightblue}-- &\cellcolor{lightblue}-- \\
B1706-44 &0.102 &74.9 &GMRT &60476 &650 &20 &6039 &1620 &0/79 &0 &-- &-- &-- \\
& & &Parkes &57309 &1382 &64 &1917 &82 &0/2 &0 &-- & & \\
& & &Parkes &55999 &732 &64 &587 &206 &0/12 &0 &-- & & \\
\cellcolor{lightblue}B1719-37 &\cellcolor{lightblue}0.236 &\cellcolor{lightblue}99.5 &\cellcolor{lightblue}Parkes &\cellcolor{lightblue}57193 &\cellcolor{lightblue}1382
&\cellcolor{lightblue}64 &\cellcolor{lightblue}832 &\cellcolor{lightblue}0 &\cellcolor{lightblue}-- &\cellcolor{lightblue}-- &\cellcolor{lightblue}-- &\cellcolor{lightblue}-- &\cellcolor{lightblue}-- \\
\cellcolor{lightblue} &\cellcolor{lightblue} &\cellcolor{lightblue} &\cellcolor{lightblue}Parkes &\cellcolor{lightblue}57165 &\cellcolor{lightblue}1382
&\cellcolor{lightblue}64 &\cellcolor{lightblue}819 &\cellcolor{lightblue}9 &\cellcolor{lightblue}0/1 &\cellcolor{lightblue}0 &\cellcolor{lightblue}-- &\cellcolor{lightblue} &\cellcolor{lightblue} \\
B1732-07 &0.419 &73.5 &GBT &60437 &820 &82 &722 &601 &0/14 &0 &-- &-- &-- \\
\cellcolor{lightblue}B1737+13 &\cellcolor{lightblue}0.803 &\cellcolor{lightblue}48.7 &\cellcolor{lightblue}GBT &\cellcolor{lightblue}60436 &\cellcolor{lightblue}820
&\cellcolor{lightblue}82 &\cellcolor{lightblue}377 &\cellcolor{lightblue}292 &\cellcolor{lightblue}0/24 &\cellcolor{lightblue}0 &\cellcolor{lightblue}-- &\cellcolor{lightblue}-- &\cellcolor{lightblue}-- \\
J1740+1000 &0.15 &23.9 &GMRT &60715 &650 &5 &3125 &398 &0/94 &0 &-- &480 $\pm$ 270 &(M) \\
& & &GMRT &60715 &650 &41 &3189 &244 &0/27 &0 &-- & &\\
\cellcolor{lightblue}B1845-19 &\cellcolor{lightblue}4.308 &\cellcolor{lightblue}18.2 &\cellcolor{lightblue}GBT &\cellcolor{lightblue}60436 &\cellcolor{lightblue}820
&\cellcolor{lightblue}82 &\cellcolor{lightblue}246 &\cellcolor{lightblue}157 &\cellcolor{lightblue}2/74 &\cellcolor{lightblue}1.3 &\cellcolor{lightblue}734$-$1700 &\cellcolor{lightblue}-- &\cellcolor{lightblue}-- \\
B1857-26 &0.612 &38.0 &GMRT &60531 &650 &20 &200 &163 &0/59 &0 &-- &-- &-- \\
\cellcolor{lightblue}B1859+03 &\cellcolor{lightblue}0.655 &\cellcolor{lightblue}401.2 &\cellcolor{lightblue}GBT &\cellcolor{lightblue}60454 &\cellcolor{lightblue}820 
&\cellcolor{lightblue}82 &\cellcolor{lightblue}461 &\cellcolor{lightblue}446 &\cellcolor{lightblue}0/20 &\cellcolor{lightblue}0 &\cellcolor{lightblue}-- &\cellcolor{lightblue}-- &\cellcolor{lightblue}-- \\
B1929+10 &0.226 &4.8 &Parkes &57981 &1182 &64 &2486 &2114 &58/285 &2.7 &760 $\pm$ 148 &480 $\pm$ 180 &(M) \\
& & &GBT &60715 &650 &64 &3189 &244 &0/27 &0 &-- & & \\
\toprule
\end{tabular}
\begin{tablenotes}[hang]
\item CF: Central Frequency; ${T_{samp}}$: Sampling Time; Obs.: Observation; $\boldsymbol{n_{pulses}}$: Number of pulses; MS: Microstructure
\item[-] References --- Li: \citet{li2025fast}; M: \citet{mitra2015polarized}; K: \citet{kramer2002high}; L: \citet{lange1998radio}.
\end{tablenotes}
\end{sidewaystable*}
\begin{figure*}[ht!]
\centering
\includegraphics[width=1\linewidth]{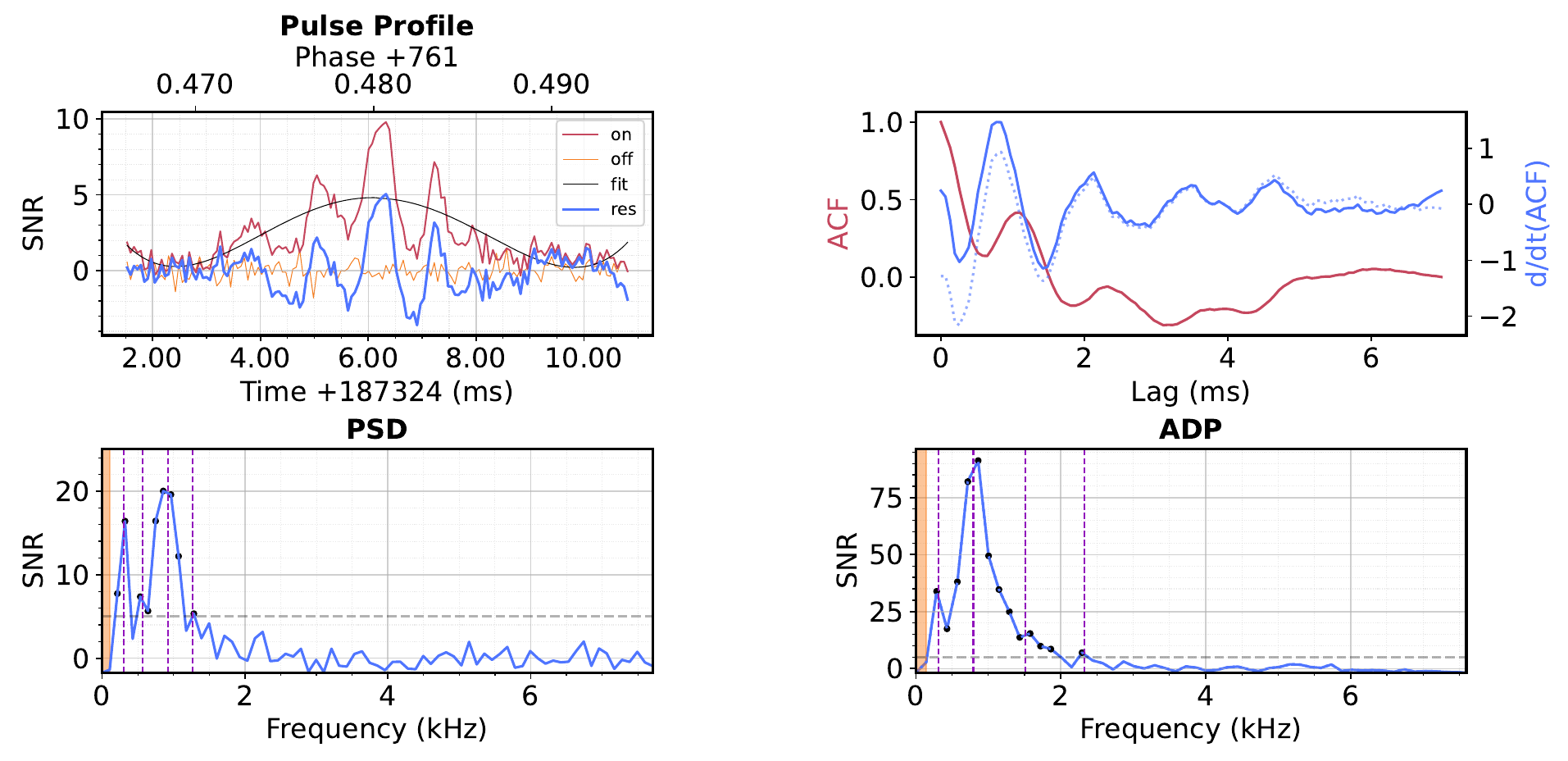}
\caption{The diagnostic plots for one of the pulses exhibiting quasi-periodic microstructures from pulsar PSR~B0540+23.}
\label{b0540result}
\end{figure*}

The sampling time of all the data used in our study, from all the three telescopes, is shorter than 256 $\mu$s. The data spans various frequencies from 550 MHz to 3 GHz. The majority of the GMRT observations were conducted at 650\,MHz with a bandwidth of 200\,MHz. All the GBT observations had a bandwidth of 200 MHz, centred at 820\,MHz. The majority of the Parkes data utilised here correspond to L-band, with bandwidths of 400\,MHz and 64\,MHz. A few of these data sets correspond to 732\,MHz and one observation was centred at 3.1\,GHz. A summary of all the data can be found in Table~\ref{tab:survey}.
RFI from the GBT and GMRT data were mitigated using RFIClean \citep{maan2021fourier}, followed by additional cleaning using \texttt{rfifind} from \texttt{PRESTO}. For the Parkes data, RFI mitigation was carried out using the \texttt{rfifind} command. The data were then folded using the \texttt{prepfold} command, typically with 512 phase bins, and a fine search in rotation period and DM was performed. The optimal DM obtained from this step was subsequently used to dedisperse the data using \texttt{prepdata} and obtain a dedispersed time-series, that was used with \qmist to search for quasi-periodic microstructures. All the above processing steps were performed in the topocentric frame.
\subsection{Results}
We performed the search for quasi-periodic microstructures using \qmist, for all the pulses with SNR$>$10, and a threshold of 5 was used in selecting the features in the $ADP_{SNR}$ and $PSD_{SNR}$. The diagnostic plots output from \qmist for all the pulses potentially containing microstructures were manually examined, and only the genuine microstructure pulses were used in any further analysis. The results are summarised in Table~\ref{tab:survey}. 
For all the pulsars with microstructure detections, we also computed the microstructure period distributions, except for pulsars for which only a few pulses with microstructures were detected. The bin width in the above histograms is computed using the Freedman-Diaconis rule, $w_{bin} = \frac{2 * IQR(x)}{n^{\frac{1}{3}}}$, where $x$ is our sample (in this case containing the microstructure periods), IQR is the inter-quartile range, and n is the size of the data. Consequently, the number of bins in the histogram is given by the integer part of $n_{bins}=\dfrac{max(x) - min(x)}{w_{bin}} + 1$. The uncertainty associated with each bin is estimated assuming Poisson statistics. The histograms are also normalised by area. 
The typical $P_{\mu}$ values listed in the eleventh column of Table~\ref{tab:survey} represent the median of the corresponding $P_{\mu}$ distributions along with the median absolute deviation (MAD). However, for B1451$-$68, B1706$–$16, and for B0540+23 at MJD~57101, the number of single pulses exhibiting microstructures is fewer than ten; therefore, the arithmetic mean and standard deviation are reported instead. In the case of B0950+08, at MJD~60442, the distribution contains only three bins, so the standard deviation was computed and then converted to MAD.
Many of the single pulses in several datasets, particularly those obtained from the Parkes archive, had SNR below our threshold ($\sim 10$), which hindered the detection of microstructures from the corresponding pulsars. Out of the 27 pulsars, we detect quasi-periodic microstructures for 9 pulsars. For 4 of these pulsars, it is the first time that quasi-periodic microstructures have been detected. In the following subsections, we provide details of our results for some of the individual pulsars, including the ones for which quasi-periodic microstructures have been detected for the first time as well as another pulsar that is known to exhibit such features but our analysis has revealed interesting features in this pulsar.

\begin{figure*}[ht!]
\centering
\includegraphics[width=1\linewidth]{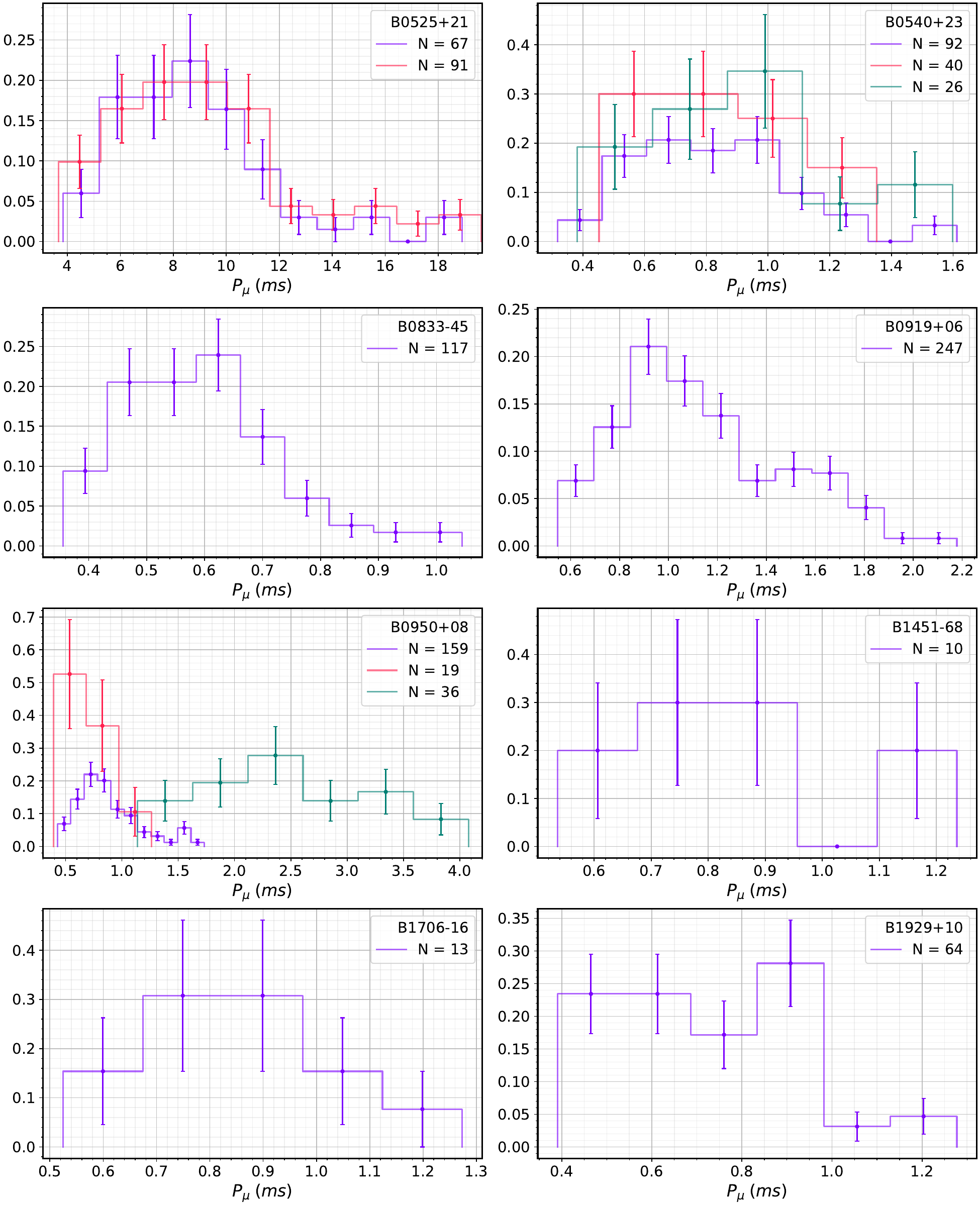}
\caption{The distributions of microstructure periods for all the pulsars in which these features were detected in this work. The distributions are shown separately for separate epochs, except for pulsars B1451$-$68 and B1706$-$16, for which pulses across all the epochs were combined. The legends show the total number of $P_{\mu}$ values included in the respective distributions, marked as ``N''.}
\label{allperiods}
\end{figure*}
\subsubsection{PSR~B0540+23}
Microstructures in B0540+23 were reported in previous studies (e.g. \citet{lange1998radio}). However, a $P_{\mu}$ has not been reported for the pulsar so far. In our analysis, we were able to determine the $P_{\mu}$ in all the four datasets we examined. All the datasets come from Parkes, and at all four epochs, more than 50\% of the single pulses had SNR greater than 10. The $P_{\mu}$ distributions as well as the median values are consistent across all epochs. The median $P_{\mu}$ for PSR~B0540+23 varies between 733\,$\mu$s and 990\,$\mu$s, but is very well consistent within the associated uncertainties. The diagnostic plots for one of the pulse components that exhibit quasi-periodic microstructures are shown in Figure~\ref{b0540result}. Two of the pulses that were found to exhibit quasi-periodic microstructures are shown in Figure~\ref{new_ms}. The $P_{\mu}$ distributions obtained for three epochs where the number of pulses detected with microstructures is more than ten are shown in Figure~\ref{allperiods}.
\subsubsection{PSR~B0950+08}
In the case of PSR~B0950+08, we observed a notable variation in the detected microstructure periodicity across different observing epochs. While one epoch (MJD~57044) yielded a typical periodicity of approximately 0.8 ms, another (MJD~57452) resulted in a significantly larger value of about 2.4 ms. Moreover, the distributions of microstructure periodicities at these two epochs overlap only a little and with clearly separated peaks, indicating a statistically significant change in the characteristic timescale of the microstructures. This discrepancy is unlikely to be attributed to instrumental effects or analysis artefacts, as both datasets were processed using the same pipeline. The absence of corresponding periodic features in the off-pulse regions further suggests that these detections are not caused by RFI. Interestingly, the fraction of single pulses exhibiting microstructure at MJD~57452 is also comparatively low.

\begin{figure*}[ht!]
\centering
\includegraphics[width=1\linewidth]{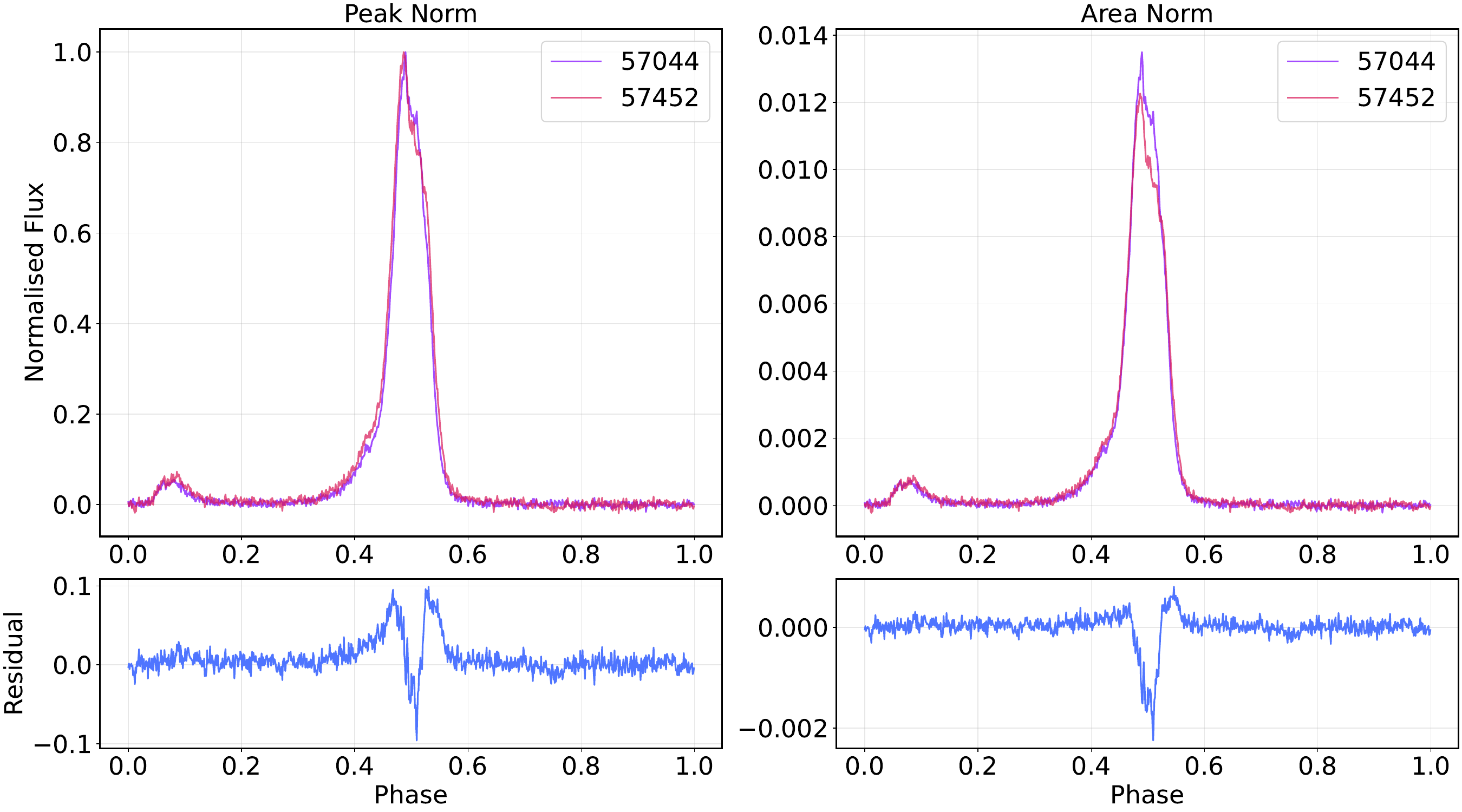}
\caption{Top: Folded pulse profiles of B0950+08, aligned in phase and normalised by their peak amplitude (left) and integrated areas (right). Bottom: Residual profile obtained by subtracting the average profile at MJD~57452 from that at MJD~57044.}
\label{fig:b0950_profile}
\end{figure*}
The change in the microstructure periodicity might indicate a change in the related emission processes. If that is indeed the case, it is worth examining any corresponding changes in the average profiles at the two epochs. With that motivation, In Figure~\ref{fig:b0950_profile}, we present the pulse profiles from epochs 57044 and 57452 overlaid on each other, and each consisting of a common number of bins (956). To investigate any significant changes in the profiles, we have first phase-aligned the profiles and then employed two ways to normalise them before obtaining a difference. In the first approach, the peaks in the individual profiles are normalised to 1, while in the second approach the individual profiles are normalised by their integrated areas. In both the approaches, the baselines were estimated and removed before the normalisation. In both the approaches, the residuals obtained by subtracting one profile from the other reveals features that cannot be attributed to noise, indicating differences in the emission properties between the two epochs, i.e., a mode-change between the two epochs. We note that the observation length is more than 2000 rotation periods at both the epochs, and the average profile is expected to reach a stable shape over such integrations. This mode change could potentially be linked to the significant change in microstructure periodicity seen above.
We note that the sampling time at MJD~57452 was 256\,$\mu$s, i.e., coarser by a factor of 4 than that at the other two epochs (64\,$\mu$s). To examine the effect of this coarser sampling time, we down-sampled the MJD~57044 data by a factor of 4 and searched for quasi-periodic microstructures. The histogram of the resulting microstructures (not shown here but assessed separately) peaked around 2.2\,ms, similar to what is seen for MJD~57452. Thus, the most likely reason for the microstructures to have a longer periodicity at MJD~57452 is the coarser sampling time, and not linked to the mode-change discussed above.
\citet{li2025fast} has reported the presence of microstructures in the interpulses (IPs) of B0950+08. In our analysis of B0950+08, we also found two IPs with microstructures at the epoch MJD~57044, though a clear periodicity is not obvious in these. Other pulsars in our sample that exhibit IPs include B1055$-$52 and J1549$-$4848. For B1055$-$52, we did not see microstructures in either the main pulse or the IPs, although there were 233 IPs above SNR=10. In the case of J1549$-$4848, the SNR of pulses were too poor to search for periodic microstructures.

\subsubsection{PSR~B1451$-$68}
All three datasets of PSR~B1451$-$68 that we have analysed come from the Parkes telescope, at 1.4\,GHz. Across the three observational epochs, quasi-periodic microstructure emission was identified at two epochs. At MJD~56866, only two pulses exhibited detectable microstructures, while at MJD~57129, eight such pulses were found. For the latter epoch, the arithmetic mean $P_{\mu}$ is provided in Table~\ref{tab:survey}. The overall mean $P_{\mu}$ estimate using all the pulses is $761\pm116$\,$\mu$s. Our findings constitute the first reported detection of quasi-periodic microstructures from this pulsar. Two of the pulses with quasi-periodic microstructures from this pulsar are shown in Figure~\ref{new_ms}.
\subsubsection{PSR~B1706$-$16}
We used three datasets of PSR~B1706$-$16 at 1.4\,GHz (see Table~\ref{tab:survey}). With a flux density of $\sim$4.1 mJy at this frequency, the resulting SNR of the individual pulses from this pulsar were typically lower than those of the other pulsars in which quasi-periodic microstructures were detected. Nevertheless, we were able to identify quasi-periodic microstructures in four single pulses from dataset at MJD~60636, and in eight single pulses at MJD~58012 (see Appendix~\ref{sec:resultplots} for a diagnostic plot). These two datasets come from two different telescopes, GMRT and Parkes. At the third epoch, MJD~57039, the detection significance of the pulsar was relatively low, potentially due to interstellar scintillation or lower sensitivity of the observation. This is the first time the detection of quasi-periodic microstructures has been made from this pulsar. Two of the pulses with quasi-periodic structures from this pulsar are shown in Figure~\ref{new_ms}. The arithmetic means $P_{\mu}$ for the two epochs are provided in Table~\ref{tab:survey}, and the overall mean $P_{\mu}$ is 866$\pm$118\,$\mu$s.
\subsubsection{PSR~B1845$-$19}
\begin{figure*}
 \centering
 
 \begin{subfigure}{1\linewidth}
 \centering
 \includegraphics[width=0.8\linewidth]{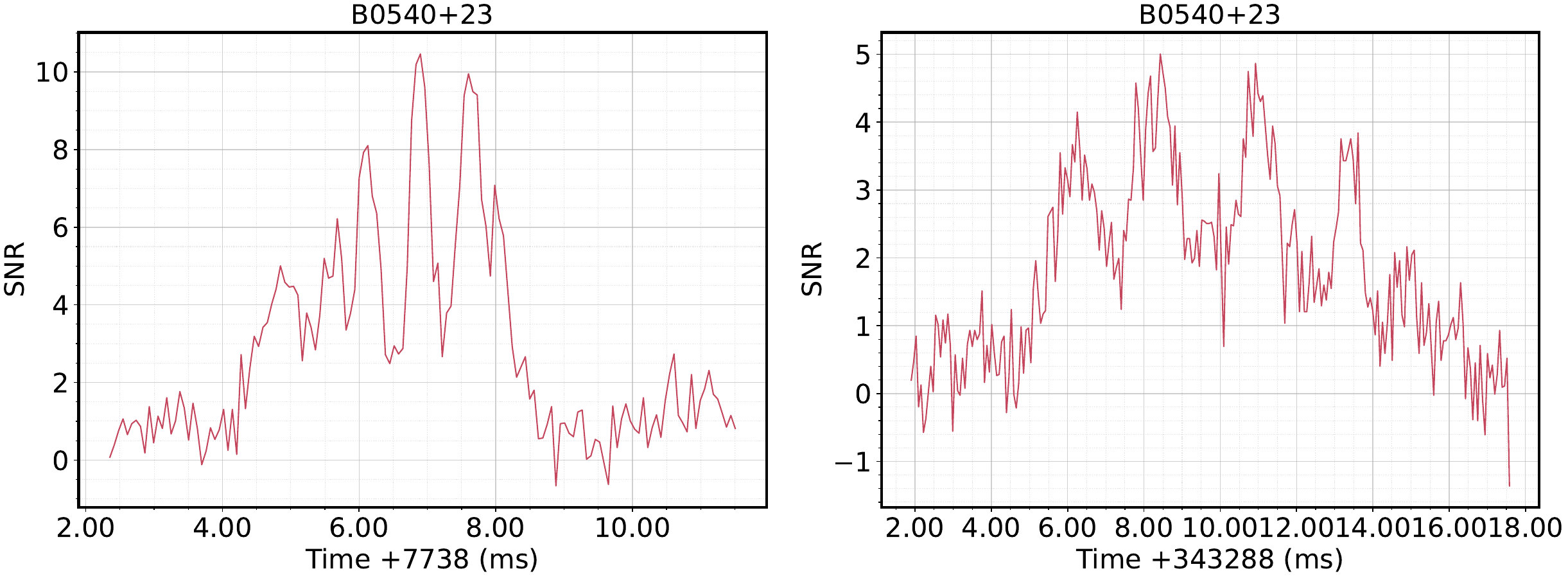}
 \end{subfigure}
 
 \vspace{\baselineskip}
 
 \begin{subfigure}{1\linewidth}
 \centering
 
 \includegraphics[width=0.8\linewidth]{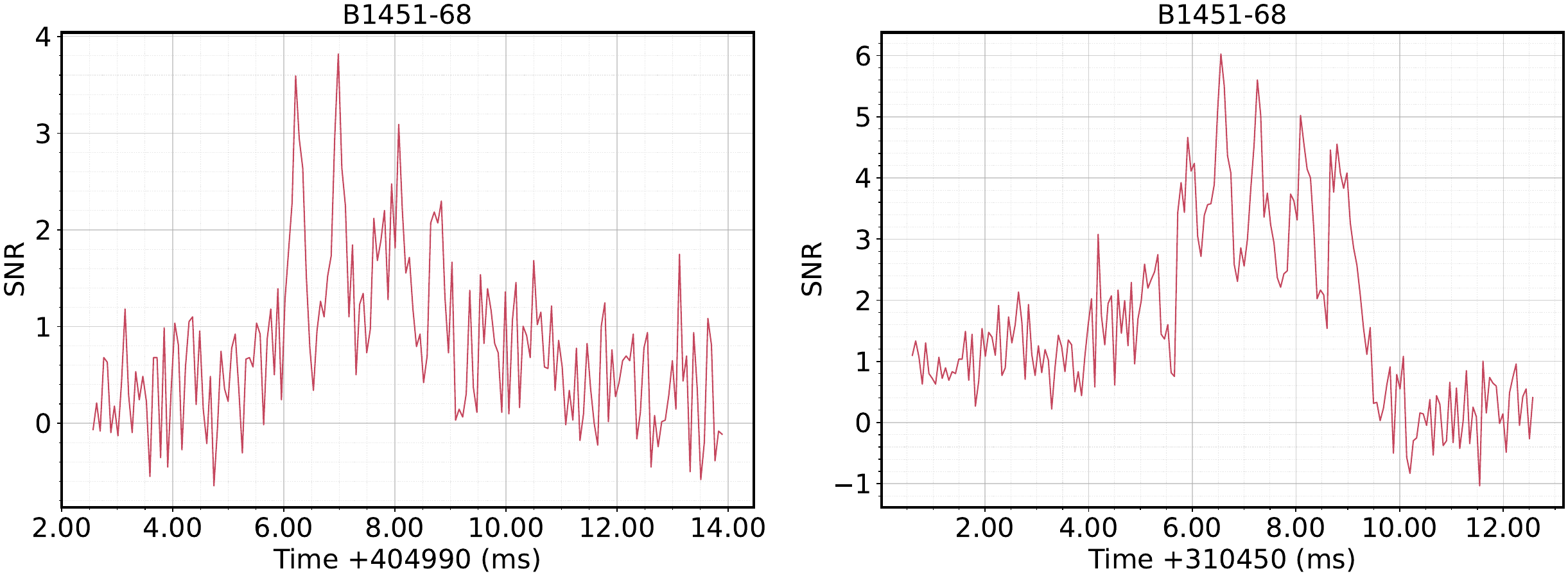}
 \end{subfigure}
 
 \vspace{\baselineskip}
 
 \begin{subfigure}{1\linewidth}
 \centering
 \includegraphics[width=0.8\linewidth]{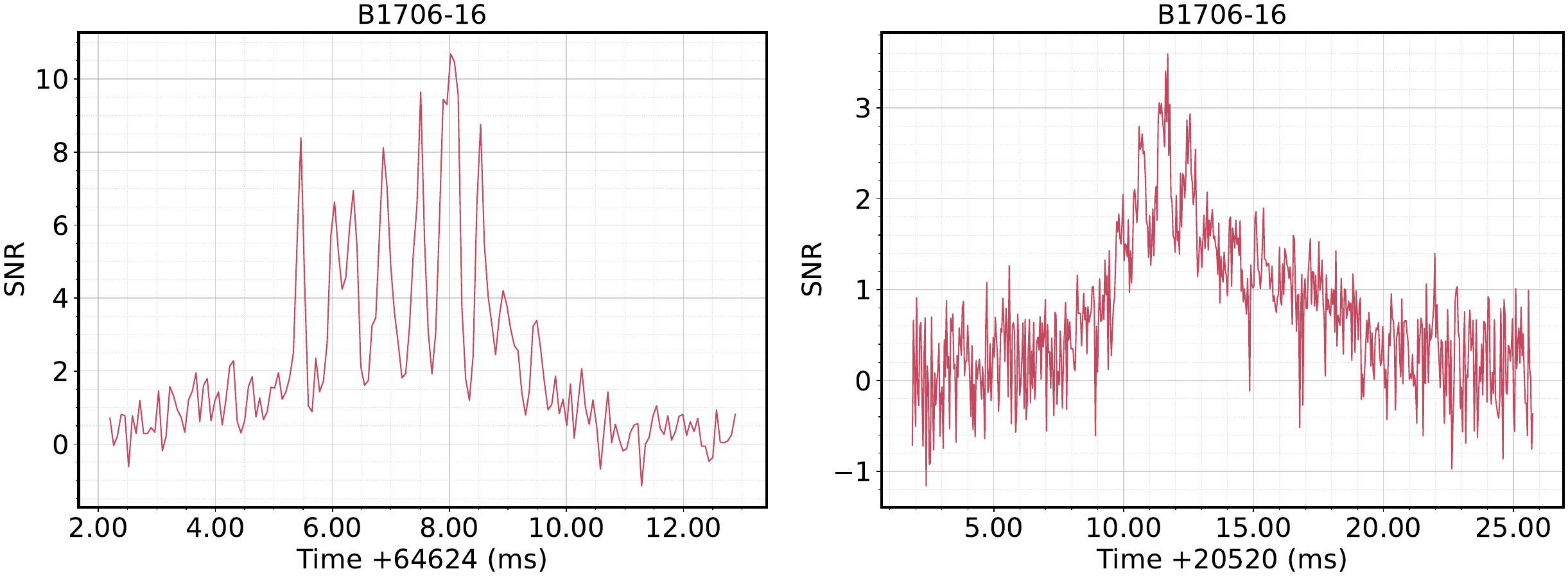}
 \end{subfigure}
 
 \vspace{\baselineskip}
 
 \begin{subfigure}{1\linewidth}
 \centering
 \includegraphics[width=0.8\linewidth]{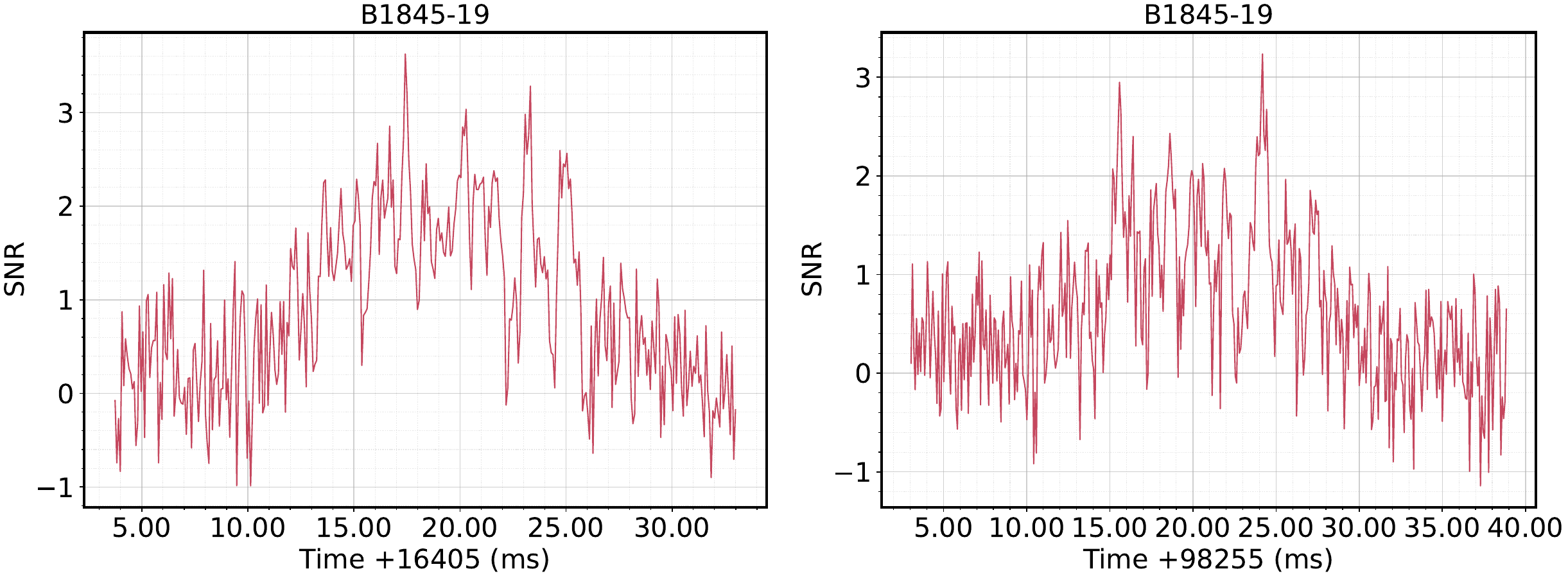}
 \end{subfigure}
 \caption{Example single pulses from the four pulsars which have been discovered to exhibit microstructure in this work. Each row of panels shows two representative pulses from one pulsar, and the four rows, from top to bottom, correspond to the pulsars B0540+23, B1451$-$68, B1706$-$16 and B1845$-$19.}
 \label{new_ms}
 \end{figure*}

We have access to a dataset from a single epoch for this pulsar. From our analysis of this dataset, two individual pulses exhibiting microstructure were identified, marking the first detection of microstructures from this pulsar. The estimated values of $P_{\mu}$ in these two pulses are 734\,$\mu$s and 1700\,$\mu$s. As only two pulses were found to exhibit microstructures, we provide both the values in Table~\ref{tab:survey} instead of a formal mean estimate. Both the pulses with quasi-periodic structures from this pulsar are shown in Figure~\ref{new_ms}.
\par
Example diagnostic plots for each of the four pulsars, B0540+23, B1451$-$68, B1706$-$16, and B1845$-$19, are presented in Appendix~\ref{sec:resultplots}.

\vspace*{4mm}
The microstructures observed in most single pulses are typically superposed on a broad emission envelope. However, for pulsars PSR~B0950+08 and PSR~B0540+23, we also identified a distinct subset of single pulses in which this underlying emission envelope is weak or altogether absent, leaving only the narrow periodic features appearing above the noise floor. Similar pulse-to-pulse morphological variations in B0950+08 have been reported previously by \citet{li2025fast}. Within this category of single pulses, sometimes there are also microstructures which are not necessarily quasi-periodic in nature.

\section{DISCUSSION}
\label{sec:discussion}
\subsection{\qmist: A partially automated solution to search for microstructures}
\qmist provides an automated and systematic framework for detecting quasi-periodic microstructures in radio pulsar emission, addressing a long-standing limitation of earlier approaches that relied heavily on manual inspection of every individual single pulse. Once a time-series is provided and the phase-range of interest is specified, \qmist automatically extracts the single pulses, estimate their widths and SNRs, computes ACF, PSD and ADP, robustly estimate the significances of peaks in the PSD and ADP, and finally outputs candidate pulses exhibiting quasi-periodic microstructures, along with the diagnostic plots and other related information. The diagnostic plots can be manually inspected to make the final decision whether the candidate pulses indeed exhibit quasi-periodic microstructures. \qmist also has the provision to analyse the off-pulse signals to investigate any periodic RFI present in the data, and hence in the identification and discarding of any spurious detections.
\par
We have successfully validated \qmist using data on pulsars that are known to exhibit quasi-periodic microstructures as well as by discovering such features from several other pulsars. The efficiency of the pipeline is visible from Table~\ref{tab:survey} --- for 55\% of the datasets that are listed there, less than 5\% of the total pulses were output as candidates for which the diagnostic plots had to be manually examined. Similarly, for 75\% of the datasets listed in Table~\ref{tab:survey}, the output fraction grows only up to 15\% of the total pulses. We note that some of the datasets in Table~\ref{tab:survey} were contaminated by periodic RFI and baseline variations, and hence resulted in increased output fraction. Of the total output candidate pulses, the fraction of the pulses with true microstructures varies up to 45\%, however, for the majority of cases it is less than 15\%, highlighting the importance of the final manual inspection. This is one place where a machine learning based automation of the decision making will further automate the overall search procedure and \qmist. We note that very recently, during the preparation of this manuscript, \citet{wang2026identifying} demonstrated the efficacy of using convolutional neural networks to detect quasi-periodic microstructures from pulsars observed with the FAST telescope. A similar approach to the diagnostic plots output from \qmist, which is agnostic to the telescope used for the observations, will be useful in the classification of the final candidates. 
\subsection{Quasi-periodic microstructure survey and new discoveries}
In Section~\ref{sec:survey}, we presented a systematic survey of quasi-periodic microstructures in a sample of young and normal pulsars, in both the northern and southern parts of the sky, using \qmist. By applying a uniform detection methodology across multi-epoch datasets obtained from different telescopes and at different frequency bands, we ensured consistency and reproducibility of the analysis. Our survey successfully recovered previously reported microstructure detections, e.g., from pulsars B0525+21, B0950+08, B1929+10, etc. Among these, for B0950+08, we observed a large difference in the mean $P_{\mu}$ at one of the epochs. In our probes to determine the origin of the discrepant $P_\mu$ at one epoch, we also discovered a mode-change event in this pulsar. However, our analysis demonstrates that the origin of the discrepant $P_\mu$ is likely in the coarse sampling time used at that particular observing epoch, and not due to the mode-change event. Thus, to obtain the intrinsic periodicity of the microstructures, it is important to have appropriately high time resolution.
The survey resulted in the discovery of quasi-periodic microstructures from four more pulsars which, to the best of our knowledge, were not known to exhibit such features so far. One of these four pulsars, B0540+23, was known to exhibit microstructures but it is for the first time that the underlying periodicity has been measured. The estimated mean $P_{\mu}$ of B0540+23 is found to be consistent across three independent observing epochs.  For the other three pulsars, B1451$-$68, B1706$-$16, and B1845$-$19, it is the first time that the microstructures and their underlying periodicity have been found. All the four pulsars are fairly old, with the youngest and oldest, B0540+23 and B1451$-$68, having characteristic ages of 0.2 and 42\,Myrs, respectively. 
Except for one pulsar, J1740+1000, we were able to detect quasi-periodic microstructures from all the pulsars in our sample which were known to exhibit such features. Given the relatively small sample size compared to the number of known pulsars and the heterogeneous nature of the datasets in our sample, in terms of sensitivity (multiple telescopes), observing frequency, observing configuration, etc., it would be difficult to conclude how much fraction of the pulsar population might exhibit the microstructure phenomena. However, the sample level statistics might still be instructive, especially as our sample was not biased against the known microstructure emitting pulsars. Including J1740+1000, 10 pulsars in our sample are now established to exhibit quasi-periodic microstructures, i.e., 37\% of the sample size. We note that, from the three new pulsars, B1451$-$68, B1706$-$16, and B1845$-$19, only a small fraction, 0.2$-$3\%, of the pulses were found to exhibit quasi-periodic microstructures. When compared to all the pulsars in our sample from which we detected microstructures, this fraction is at the lowest end. This finding suggests that many pulsars, especially the ones currently not known to exhibit microstructures, might have such features only in a small fraction of their single pulses. For a few of the pulsars in our sample (e.g., J0407+1607), the sampling time might also have not been adequately fine. The fact that the datasets for many pulsars in our sample comprise of only a few hundred pulses (in one case less than a hundred!) might also have limited the survey yield. Furthermore, in some cases, particularly for fainter pulsars and observations with limited sensitivity, only a small fraction of the single pulses were detectable, which might have imposed further limitations in detecting microstructures. Hence, the above fraction of 37\% is strictly a lower limit, and the actual fraction of pulsars exhibiting microstructure is likely to be much larger. A survey consisting of sufficiently large sample of pulsars, with adequately high time resolution and high sensitivity for all the sources in the sample, will be useful in not only determining the above fraction more meaningfully but also in probing any dependence of microstructure emission on the physical parameters of the pulsars.
\begin{figure*}[ht!]
\centering
\includegraphics[width=0.9\linewidth]{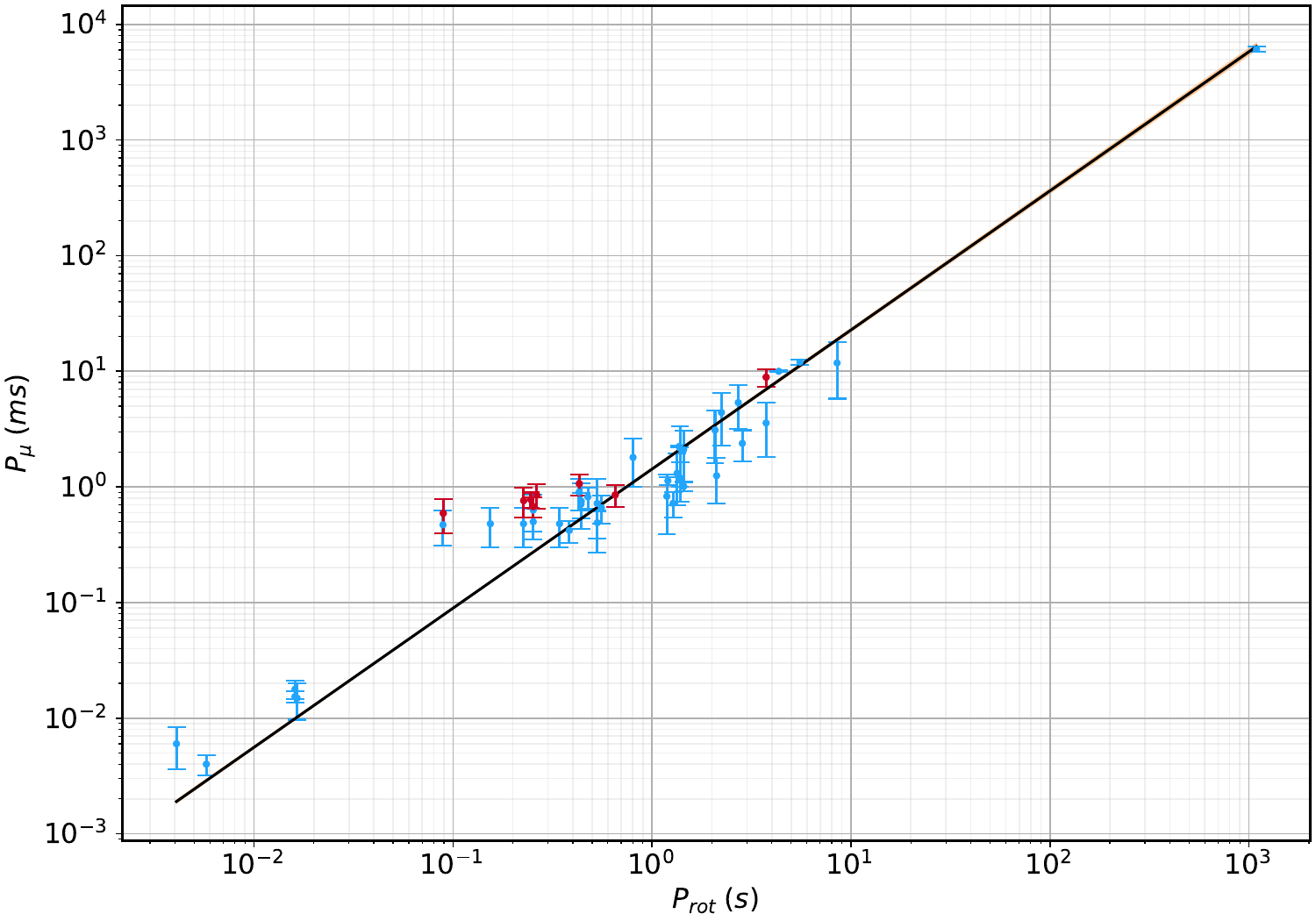}
\caption{Measured $P_{\mu}$ for various pulsars, including those from literature (blue points) as well as from this work (red points), are shown as a function of the rotation period. The fitted model (Equation ~\ref{fit_eq}) is shown as the black line along with the shaded uncertainty region.}
\label{Prot}
\end{figure*}
\subsection{$P_{rot}$\mdash$P_{\mu}$ relationship and origin of microstructures}
The relationship between $P_{rot}$ and $P_{\mu}$ in different kinds of neutron stars (pulsars, magnetars and RRATs) has been investigated earlier \citep[e.g.,][]{li2025fast,kramer2024quasi,de2016detection}. We compiled the mean $P_{\mu}$ values reported in previous studies, together with those derived from our survey, including the new discoveries (in case of detections at multiple epochs, we use a weighted mean, see Table in \ref{sec:puprot}), and used them to estimate the parameters of the power-law function: 
\begin{equation}
     P_{\mu} = A \times P_{rot}^{\alpha}
\end{equation}
Rather than applying a least-squares based fit directly to this function, we first linearise the relation in the logarithmic space:
\begin{equation}
    log(P_{\mu}) = log(A) + \alpha log(P_{rot})
\end{equation}
We then express the relation in base-10 exponential form.
\begin{equation}
    P_{\mu} = 10^{a + \alpha log(P_{rot})}
    \label{fit_eq}
\end{equation}
where $a = log(A)$. This formulation preserves the approximate Gaussian nature of the uncertainties in $P_{\mu}$ in logarithmic space and reduces bias during the fitting procedure.
The parameters $a$ and $\alpha$ and their associated uncertainties were estimated using a Markov Chain Monte Carlo (MCMC) framework. Uniform priors in the range $[-10, 10]$ were adopted for both the parameters. Initial parameter values were obtained by fitting Equation~\ref{fit_eq} using \texttt{curve\_fit} function from the \texttt{SciPy} library. The MCMC sampling was then initialised around these values using a Gaussian distribution with 64 walkers and was run for 10,000 iterations.
The log-likelihood function was defined as
\begin{equation}
\ln p(y \mid \text{model}) = -\frac{1}{2} \sum \left[ \frac{(y_{\text{data}} - y_{\text{model}})^2}{\sigma^2} + \ln(2\pi \sigma^2) \right]
\label{eq:loglikelihood}
\end{equation}
Final estimates for the parameters $a$ and $\alpha$ were obtained from the median (50th percentile) of the marginal posterior distributions. The corresponding $A$ value was computed as $A = 10^{a}$. Uncertainties were quantified using the 16$^{th}$ and 84$^{th}$ percentiles of the posterior distributions, corresponding to approximately $1\sigma$ confidence intervals. The corner plot (now shown here but assessed separately) indicates that the fitted parameters are well constrained.
As shown in Figure \ref{Prot}, we obtain the following constrained relation between $P_{rot}$ and $P_{\mu}$:
\begin{equation}
    P_{\mu} (ms)= (1.43 \pm 0.08) \times P_{rot}(s)^{1.20 \pm 0.01}
    \label{prot_eq}
\end{equation}
Thus, our results reconfirm the empirical, near-linear relationship between the pulsar rotation period, $P_{rot}$ and the microstructure period, $P_{\mu}$. The newly discovered $P_{\mu}$ for four pulsars in this work remain consistent with the above empirical relationship, further strengthening the evidence that the characteristic timescale of microstructures is closely linked to the fundamental rotational properties of the neutron stars.
Various $P_{\mu}$ measurements as well as the above fitted relationship are shown in Figure~\ref{Prot}. While the majority of the $P_{\mu}$ measurements follow the above fitted relationship very well, we note a curious deviating trend around rotation periods of 0.1$-$0.2\,s. The previous measurements from literature as well as our measurements from this work tend to deviate from the above linear trend in this period range. Given the limited number of $P_\mu$ measurements in this period range, detection of microstructures from more pulsars with periods around 0.1\,s will be required to draw any firm conclusions.
A few different classes of models have been proposed to explain microstructures, each attributing the quasi-periodic structures to different physical mechanisms within the pulsar magnetosphere. The geometric model \citep{benford1977model} interprets microstructures to be caused by narrow ``beamlets'' of radio emission originating from streams of relativistic particles along the curved magnetic field lines. The micropulse widths correspond to the angular size of the emission beamlets and the separation between the beamlets decides the period. In this model, the empirical scaling between $P_{\mu}$ and $P_{rot}$ is naturally explained as the microstructure timescale is coupled with the angular structure of the emission region. A temporal or radial origin of the microstructures is attributed to a radial modulation of the emission region. The radial modulation could be caused by variability in the plasma production or by the propagation effects within the magnetosphere. While this model does not naturally predict the observed relationship between $P_{\mu}$ and $P_{rot}$, it remains plausible. More recently, \citet{thompson2022radio} has proposed that beamed curvature radiation by solitons stemming from small-scale localised current sheets can naturally produce the observed microstructures. This model also predicts a geometric origin, and hence, the relationship between $P_{\mu}$ and $P_{rot}$ is naturally explained. Broadband or multi-frequency characterisation of quasi-periodic microstructures of well known emitters and characterisation of the microstructure emitting fraction of the pulsar population and any dependence on their fundamental properties using a large-scale, high-sensitivity survey of pulsars will be useful in providing further clues to the actual physical emission mechanism of these intriguing features.
\section{CONCLUSION}\label{sec:conclusion}
We have developed a software pipeline, \qmist\footnote{Publicly available at \url{https://github.com/4marnath/qmist}.}, for detecting quasi-periodic microstructures in radio pulsar emission, addressing the limitations of earlier approaches that relied heavily on manual extraction and inspection of individual pulses. Along with a detailed description of \qmist, we have also demonstrated its effectiveness in detecting microstructures. We have also discussed the possible areas where further improvements will be useful. One such area is a machine-learning based vetting of the \qmist-output diagnostic plots for candidate quasi-periodic microstructures, instead of the current manual examination.
Using \qmist on archival data from multiple telescopes, we have conducted a microstructure survey in a sample of 27 pulsars. From this survey, we have discovered quasi-periodic microstructures from four pulsars which were either not known to exhibit such features or the underlying periodicity was not known. Three of these pulsars exhibit microstructures in a relatively low fraction of pulses, indicating future microstructure surveys will benefit from long observations. All the new discoveries follow the power-law relationship between $P_\mu$ and $P_{rot}$, confirming that the microstructure emission mechanism is strongly related to the rotation period. Around rotation periods of 0.1\,s, we note a curious deviation from the above $P_\mu$\mdash$P_{rot}$. However, a large-scale survey of microstructure emission in pulsars is required to statistically confirm or rule out this deviation. Furthermore, a broadband study of quasi-period microstructures could help in distinguishing between the temporal or radial and geometric models of microstructure emission.
\begin{acknowledgements} 
This work used data from our own observations using GMRT and GBT. We thank the staff of the GMRT that made these observations possible. GMRT is run by the National Centre for Radio Astrophysics of the Tata Institute of Fundamental Research. The Green Bank Observatory is a facility of the National Science Foundation operated under a cooperative agreement by Associated Universities, Inc. This paper also includes archived data obtained through the Parkes Pulsar Data archive on the CSIRO Data Access Portal (\url{http://data.csiro.au}). Murriyang, CSIRO’s Parkes radio telescope, is part of the Australia Telescope National Facility (\url{https://ror.org/05qajvd42}) which is funded by the Australian Government for operation as a National Facility managed by CSIRO.
\end{acknowledgements}

\paragraph{Funding Statement}
YM acknowledges support from the Department of Science and Technology via the Science and Engineering Research Board Startup Research Grant (SRG/2023/002657), and for funding support under project 12$-$R\&D$-$TFR$-$5.02$-$0700.
\paragraph{Competing Interests}
None.
\paragraph{Data Availability Statement}
The data used in Figure~\ref{Prot} are listed in Appendix~\ref{sec:puprot}.
\bibliographystyle{apj}
\bibliography{refs}
\onecolumn
\appendix
\section{Relationship Between $\sigma_t$ and $\sigma_{\nu}$}
\label{sec:sigsig}
Consider a Gaussian curve in the time domain
\begin{equation}
    f(t) = \frac{1}{\sqrt{2 \pi \sigma_t^2}} 
e^{-\frac{1}{2} \left(\frac{t}{\sigma_t}\right)^2}
\label{eq:gaussian}
\end{equation}
Taking the Fourier Transform,
\begin{align*}
\mathcal{F}(f(t))(\nu) &= 
\frac{1}{\sqrt{2 \pi \sigma_t^2}}
\int_{-\infty}^{\infty} 
e^{-\frac{1}{2}\left(\frac{t}{\sigma_t}\right)^2} 
e^{-2 \pi i \nu t} \, dt \\[6pt]
&= 
\frac{1}{\sqrt{2 \pi \sigma_t^2}}
\int_{-\infty}^{\infty}
e^{-\frac{t^2}{2\sigma_t^2}}
\bigl(\cos(2 \pi \nu t) - i \sin(2 \pi \nu t)\bigr) \, dt \\[6pt]
&= 
\frac{1}{\sqrt{2 \pi \sigma_t^2}}
\Biggl[
\int_{-\infty}^{\infty} e^{-\frac{t^2}{2\sigma_t^2}} \cos(2 \pi \nu t) \, dt
- i \int_{-\infty}^{\infty} e^{-\frac{t^2}{2\sigma_t^2}} \sin(2 \pi \nu t) \, dt
\Biggr]
\end{align*}
\noindent
For the first term, we now use the identity (for \(a > 0\)) \citep{abramowitz1972handbook}:
\[
\int_{\infty}^{\infty} e^{-a t^2} \cos(2 \pi \nu t) \, dt
=\sqrt{\frac{\pi}{a}} \, e^{-\frac{\pi^2 \nu^2}{a}}
\]
\noindent
Here, \(a = \frac{1}{2 \sigma_t^2}\), and the second term is an odd function which simplifies to zero, so

\begin{equation}
\mathcal{F}(f(t))(\nu)
= 
\frac{1}{\sqrt{2 \pi \sigma_t^2}} \cdot
\sqrt{2 \pi \sigma_t^2} \,
e^{-2 \pi^2 \nu^2 \sigma_t^2}
- 0
\end{equation}
\begin{equation}
\mathcal{F}(f(t))(\nu) = e^{-2 \pi^2 \nu^2 \sigma_t^2}
\label{eq:ffttime}
\end{equation}

\noindent
Now let's take a general Gaussian in the frequency domain:
\[
g(\nu)=\frac{1}{\sqrt{2\pi}\,\sigma_\nu} \,
e^{-\frac{\nu^2}{2\sigma_\nu^2}}
\]
\noindent
Since this is equivalent to Equation \eqref{eq:ffttime}, we can equate them as
\[
e^{-2\pi^2\sigma_t^2\,\nu^2}
=\frac{1}{\sqrt{2\pi} \sigma_\nu} \quad
e^{-\frac{\nu^2}{2\sigma_\nu^2}}
\]
\noindent
Comparing the Gaussian-shaped part by equating coefficients of $\nu^2$ in the exponents gives
\[
2\pi^2\sigma_t^2\,\nu^2 \;=\; \frac{\nu^2}{2\sigma_\nu^2}.
\]
\[
2\pi^2\sigma_t^2 \;=\; \frac{1}{2\sigma_\nu^2}
\quad\Longrightarrow\quad
\sigma_\nu^2 \;=\; \frac{1}{4\pi^2\sigma_t^2}
\]
\[
\quad\Longrightarrow\quad
\;\sigma_\nu \;=\; \frac{1}{2\pi\,\sigma_t}\
\]
\newpage
\section{Example diagnostic plots}\label{sec:resultplots}
\begin{figure*}[h]
\centering
\begin{subfigure}{1\linewidth}
\centering
\includegraphics[width=0.75\linewidth]{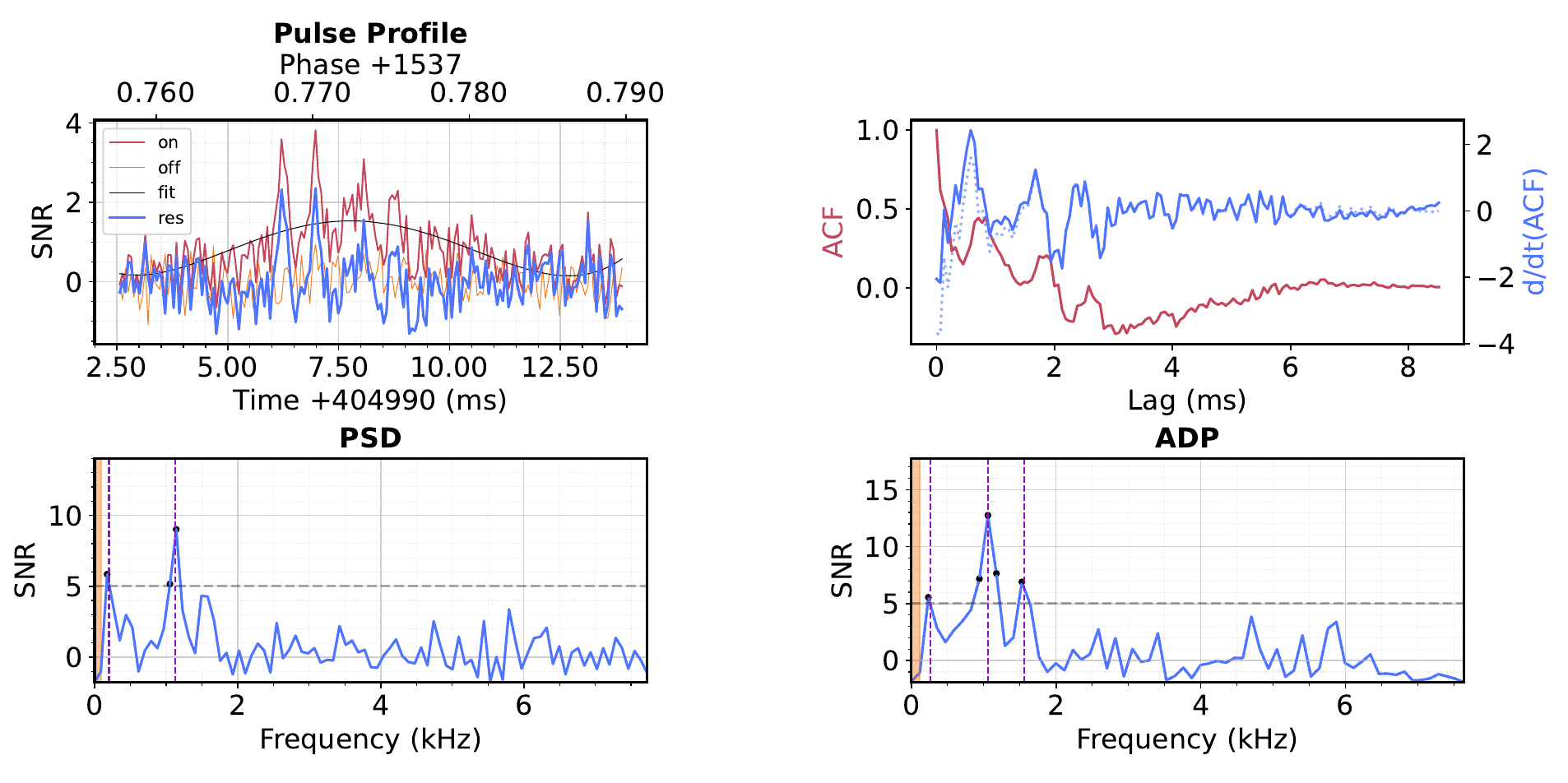}
\end{subfigure}
\begin{subfigure}{1\linewidth}
\centering
\includegraphics[width=0.75\linewidth]{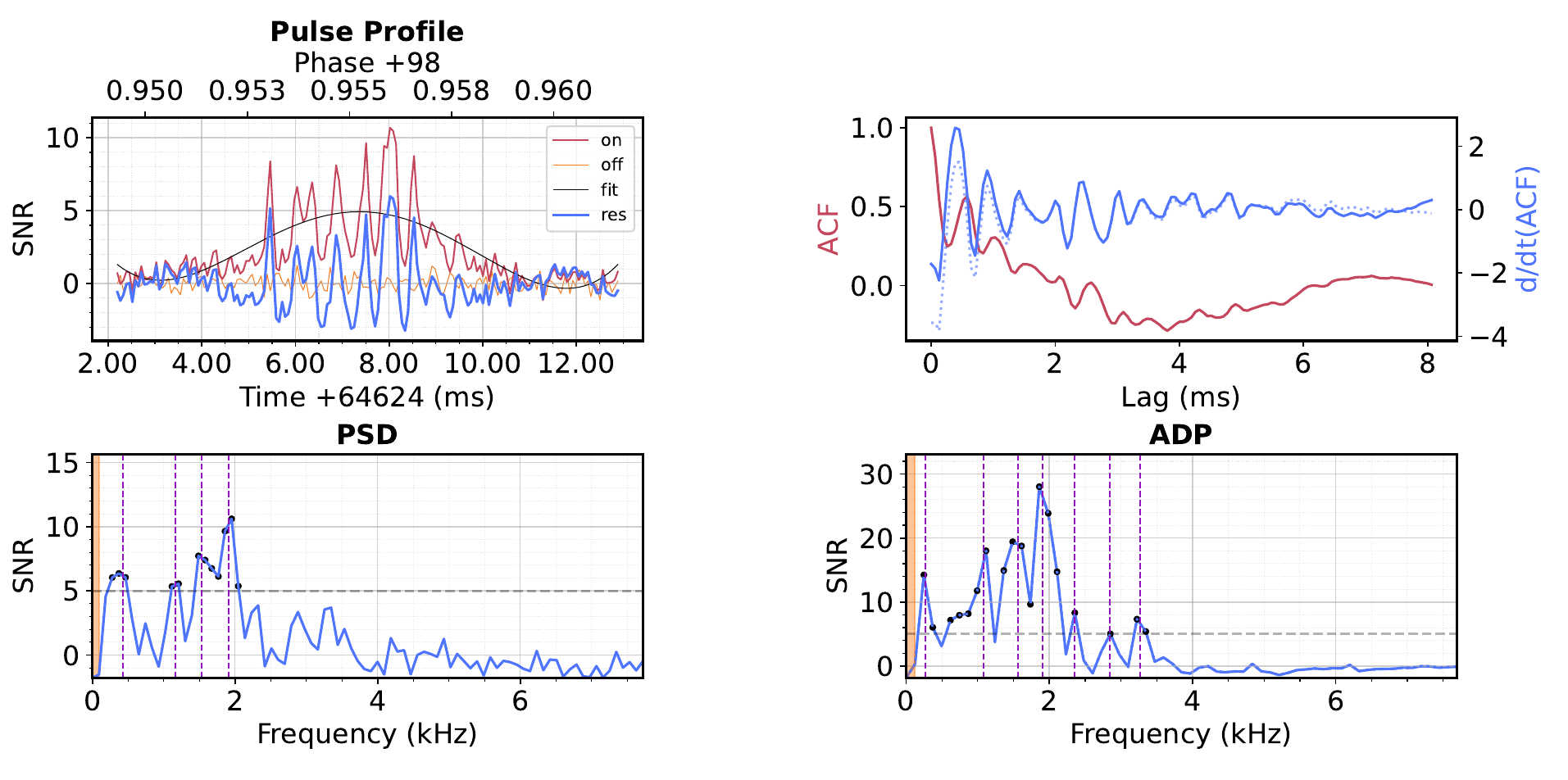}
\end{subfigure}
\begin{subfigure}{1\linewidth}
\centering
\includegraphics[width=0.75\linewidth]{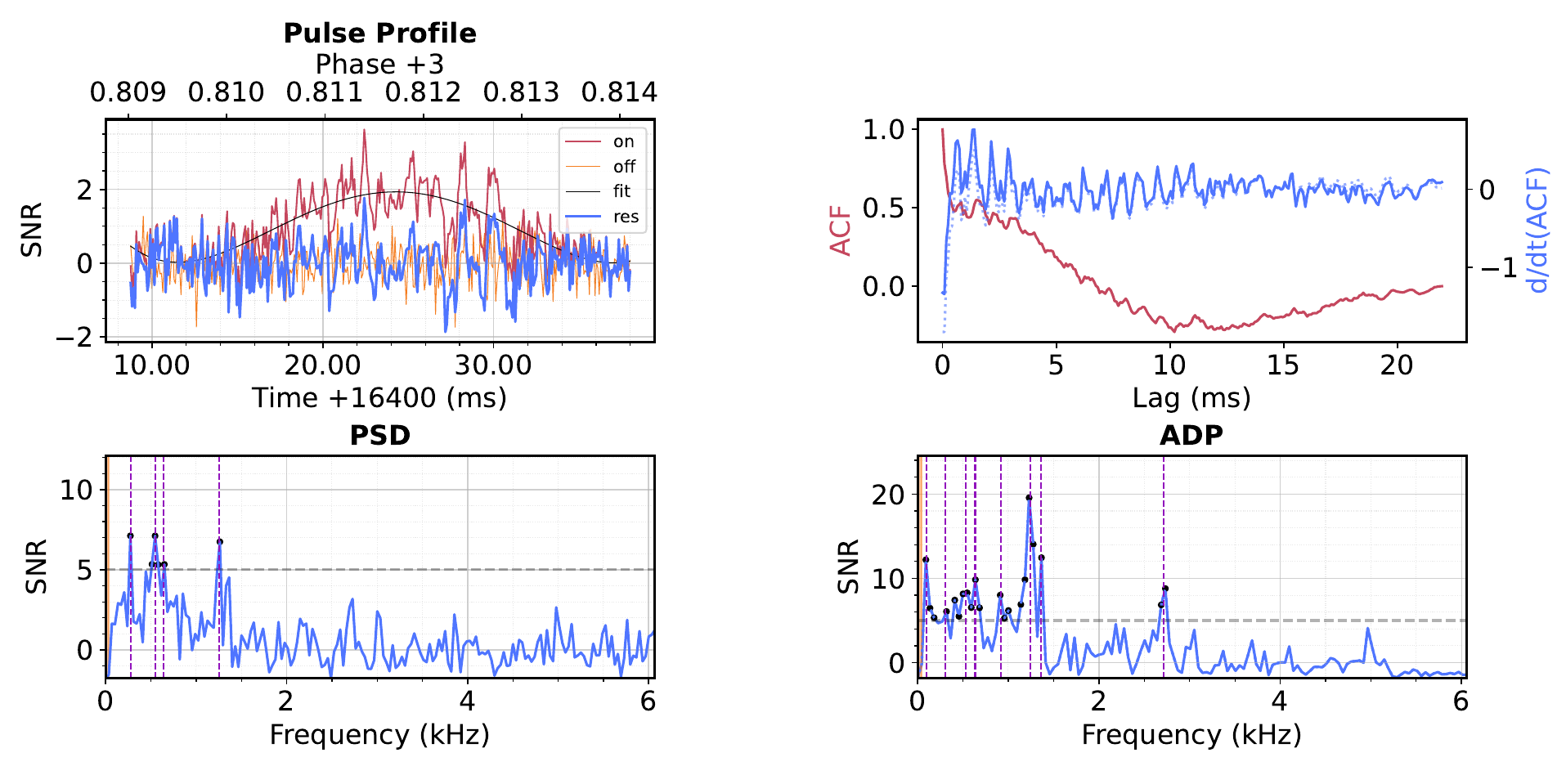}
\end{subfigure}
\caption{Example diagnostic plots for a pulse component from each of the three pulsars, B1451$-$68 (top set of plots), B1706$-$16 (middle set of plots), and B1845$-$19 (bottom set of plots), showing the detection of quasi-periodic microstructures.}
\label{newpsr_diag}
\end{figure*}

For each of the three pulsars, B1451$-$68, B1706$-$16, and B1845$-$19, for which the quasi-periodic microstructures have been found for the first time, an example diagnostic plot output from \qmist is shown here (Figure~\ref{newpsr_diag}).
\clearpage
\section{Data for Fitting $P_{\mu}$ and $P_{\rm rot}$}
\label{sec:puprot}
\begin{ThreePartTable}
\begin{longtable}{@{\extracolsep{\fill}}ccccc}
\label{tab:prot}\\
\toprule
\textbf{Pulsar Name} & \textbf{$P_{rot}$ (s)} & \textbf{$P_{\mu}$ (ms)} & \textbf{Uncertainty} & \textbf{Reference} \\
\toprule
\endfirsthead
\toprule
\textbf{Pulsar Name} & \textbf{$P_{rot}$ (s)} & \textbf{$P_{\mu}$ (ms)} & \textbf{Uncertainty} & \textbf{References} \\
\toprule
\endhead
\toprule
\endfoot
J0627+0706 &0.476 &0.82 &0.17 &Li2025 \\
J0826+2637 &0.531 &0.49 &0.13 &Li2025 \\
J0953+0755 &0.253 &0.5 &0.15 &Li2025 \\
J1946+1905 &0.441 &0.75 &0.32 &Li2025 \\
J1622-4950 &4.326 &10 &0.2 &Kramer2024 \\
J1810-197 &5.54 &12 &0.6 &Kramer2024 \\
J1818.0-1607 &1.364 &2.24 &0.04 &Kramer2024 \\
GLEAM-X &1090.8 &6110 &290 &Kramer2024 \\
J1022+1001 &0.016453 &0.0149 &0.0052 &Liu2022 \\
J2145-0750 &0.016052 &0.0178 &0.0032 &Liu2022 \\
J1744-1134 &0.004075 &0.006 &0.0024 &Liu2022 \\
J2144-3933 &8.51 &11.8 &6 &Mitra2020 \\
J0437-4715 &0.005757 &0.004 &0.00079 &De2016 \\
J2145-0750 &0.016052 &0.01536 &0.00177 &De2016 \\
B0301+19 &1.387 &1.19 &0.44 &Mitra2015 \\
B0525+21 &3.745 &3.57 &1.76 &Mitra2015 \\
J0546+2441 &2.843 &2.38 &0.71 &Mitra2015 \\
B0656+14 &0.384 &0.42 &0.09 &Mitra2015 \\
B0751+32 &1.442 &2.1 &0.98 &Mitra2015 \\
B0823+26 &0.53 &0.72 &0.45 &Mitra2015 \\
B0834+06 &1.275 &0.72 &0.18 &Mitra2015 \\
B0919+06 &0.43 &0.9 &0.27 &Mitra2015 \\
B0950+08 &0.253 &0.63 &0.22 &Mitra2015 \\
B1133+16 &1.187 &0.83 &0.44 &Mitra2015 \\
B1237+25 &1.382 &2.16 &1.16 &Mitra2015 \\
J1740+1000 &0.154 &0.48 &0.18 &Mitra2015 \\
B1737+13 &0.803 &1.8 &0.8 &Mitra2015 \\
J1910+0714 &2.712 &5.36 &2.21 &Mitra2015 \\
B1910+20 &2.233 &4.4 &2.12 &Mitra2015 \\
B1919+21 &1.337 &1.31 &0.62 &Mitra2015 \\
B1929+10 &0.226 &0.48 &0.18 &Mitra2015 \\
B1944+17 &0.44 &0.71 &0.18 &Mitra2015 \\
B2002+31 &2.111 &1.25 &0.53 &Mitra2015 \\
B2016+28 &0.557 &0.66 &0.18 &Mitra2015 \\
B2020+28 &0.343 &0.48 &0.18 &Mitra2015 \\
B2034+19 &2.074 &3.1 &1.5 &Mitra2015 \\
B2110+27 &1.202 &1.13 &0.09 &Mitra2015 \\
B2315+21 &1.444 &1.01 &0.09 &Mitra2015 \\
B0833-45 &0.089 &0.47 &0.16 &Kramer2002 \\
B0525+21 &3.7455 &8.89513 &1.54115 &This Work \\
B0540+23 &0.2459 &0.77839 &0.11602 &This Work \\
B0833-45 &0.0893 &0.59223 &0.19693 &This Work \\
B0919+06 &0.431 &1.06623 &0.21981 &This Work \\
B0950+08 &0.253 &0.67623 &0.13349 &This Work \\
B1451-68 &0.2634 &0.84734 &0.20190 &This Work\\
B1706-16 &0.653 &0.85192 &0.18684 &This Work\\
B1929+10 &0.2265 &0.76030 &0.21904 &This Work \\
\end{longtable}
\begin{tablenotes}[hang]
\item[-] Reference: Li2025: \citet{li2025fast}; Kramer2024: \citet{kramer2024quasi}; Liu2022: \citet{liu2022detection}; De2016 \citet{de2016detection}; Mitra2016: \citet{mitra2015polarized}; Kramer2002: \citet{kramer2002high}
\end{tablenotes}
\end{ThreePartTable}
\end{document}